\documentclass[aps,prd,twocolumn,showpacs,preprintnumbers,amsmath,amssymb,floatfix,nofootinbib]{revtex4-2}  

\usepackage[dvips]{graphicx}
\usepackage{latexsym}
\usepackage{amsmath}
\usepackage{epsfig}
\usepackage[T1]{fontenc} 
\usepackage{psfrag}    
\usepackage{slashed}   
\usepackage{mathrsfs}
\usepackage{latexsym,bm}
\usepackage{amsmath}
\usepackage{amssymb}

\newcommand{\bea}{\begin{eqnarray}} \newcommand{\eea}{\end{eqnarray}}

\def\Comment#1{}

\newcommand{\bean}{\begin{eqnarray*}}
\newcommand{\eean}{\end{eqnarray*}}

\newcommand{\gapproxeq}{\lower
.7ex\hbox{$\;\stackrel{\textstyle >}{\sim}\;$}}
\newcommand{\lapproxeq}{\lower
.7ex\hbox{$\;\stackrel{\textstyle <}{\sim}\;$}}

\newcommand\lsim{\mathrel{\rlap{\lower4pt\hbox{\hskip1pt$\sim$}}
    \raise1pt\hbox{$<$}}}
\newcommand\gsim{\mathrel{\rlap{\lower4pt\hbox{\hskip1pt$\sim$}}
    \raise1pt\hbox{$>$}}}
\newcommand{\ba}{\begin{array}}
\newcommand{\ea}{\end{array}}
\newcommand{\nn}{\nonumber}

\newcommand{\be}{\begin{equation}}
\newcommand{\ee}{\end{equation}}
\newcommand{\bear}{\begin{eqnarray}}
\newcommand{\eear}{\end{eqnarray}}

\newcommand{\ket}{\,\rangle}
\newcommand{\bra}{\langle \,}
\newcommand{\eqn}[1]{(\ref{#1})}
\newcommand{\cO}{{\cal O}}

\newcommand{\mF}{\mathcal{F}}

\newcommand{\mJ}{\mathcal{J}}

\newcommand{\mO}{\mathcal{O}}

\newcommand{\mT}{\mathcal{T}}

\newcommand{\mY}{\mathcal{Y}}
\newcommand{\Frac}[2]{\frac{\displaystyle #1}{\displaystyle #2}}
\newcommand{\Int}{\displaystyle{\int}}

\def\bat{\begin{array}{cc}}

\usepackage{color} 

\definecolor{mygreen}{RGB}{0,128,0} 
\definecolor{mybrown}{RGB}{153,102,51}
 
\begin{document}

\preprint{IPARCOS-UCM-25-014} 

\title{Oblique parameters at next-to-leading order within \\ electroweak strongly-coupled scenarios: constraining heavy resonances}

\author{Antonio Pich${}^{1}$}
\author{Ignasi Rosell${}^{2}$}
\author{Juan Jos\'e Sanz-Cillero${}^{3}$}

\affiliation{${}^1$ IFIC, Universitat de Val\`encia -- CSIC, Apt. Correus 22085, E-46071 Val\`encia, Spain }

\affiliation{${}^2$  Departamento de Matem\'aticas, F\'\i sica y Ciencias Tecnol\' ogicas, Universidad Cardenal Herrera-CEU, CEU Universities, E-46115 Alfara del Patriarca, Val\`encia, Spain}

\affiliation{${}^3$  Departamento de F\'\i sica Te\'orica and Instituto de F\'\i sica  de  Part\'\i culas  y  del  Cosmos IPARCOS,  Universidad Complutense de Madrid, E-28040 Madrid, Spain}

\begin{abstract} 
Using a general (non-linear) effective field theory description of the Standard Model electroweak symmetry breaking, we analyse the impact on the electroweak oblique parameters of hypothetical heavy resonance states strongly coupled to the SM particles. We present a next-to-leading order calculation of $S$ and $T$ that updates and generalizes our previous results, including P--odd operators in the Lagrangian, fermionic cuts and the current experimental bounds. We demonstrate that in any strongly-coupled underlying theory where the two Weinberg Sum Rules are satisfied, as happens in asymptotically free gauge theories, the masses of the heavy vector and axial-vector states must be heavier than 10 TeV. Lighter resonances with masses around 2-3 TeV are only possible in theoretical scenarios where the 2nd Weinberg Sum Rule is not fulfilled.
\end{abstract}



\maketitle

\vspace{-5cm}

\section{Introduction} \label{sec:introduction}

The first two runs of the LHC have confirmed the Standard Model (SM) as the right theory of the electroweak interactions at the energy scales explored so far. The discovery of a Higgs-like\footnote{Although it might not be the SM Higgs boson, we will refer to this particle as ``Higgs''.} particle~\cite{higgs}, with couplings fully compatible with the SM expectations, has completed the SM spectrum of fundamental fields and no new states have yet been observed. Therefore, the available data suggest the existence of a mass gap between the SM and any hypothetical New-Physics (NP) degrees of freedom. This gap justifies the use of effective field theories to search for fingerprints of heavy scales at low energies in a systematic way.

The main ingredients for the construction of any effective field theory are the particle content, the symmetries and the power counting. In the electroweak case, the power counting to be used depends on the way of introducing the Higgs field~\cite{Buchalla:2016bse,Pich:2018ltt}. One can consider the more common linear realization of the electroweak symmetry breaking (EWSB) \cite{Brivio:2017vri}, assuming the Higgs to be part of a doublet together with the three electroweak (EW) Goldstones, as in the SM, or the more general non-linear realization \cite{Pich:2018ltt}, without assuming any specific relation between the Higgs $h$ and the three Goldstone fields $\vec{\varphi}$. We follow here the second option~\cite{lagrangian,lagrangian_color}, where an expansion in generalized momenta is adopted. Note that the linear realization is a particular case of the more general non-linear one.

In addition to the (non-linear) electroweak effective theory, containing only the SM particle content, we consider an underlying strongly-coupled scenario incorporating   
bosonic heavy resonances with $J^P=0^\pm$ and $J^P=1^\pm$ which interact with the SM particles. In previous works, we have investigated the contributions of these heavy states to the low-energy  couplings (LECs) of the effective electroweak Lagrangian \cite{lagrangian,lagrangian_color,Pich:2015kwa,PRD2}. Here, we are going to analyse the constraints on the heavy resonance masses emerging from the electroweak oblique parameters~\cite{Peskin_Takeuchi}.

In Ref.~\cite{ST}, we already presented a one-loop calculation of the $S$ and $T$ parameters within this strongly-coupled scenario, incorporating the recently discovered Higgs boson into our previous Higgsless calculation \cite{S_Higgsless}.
Our analysis provided generic (model-independent) and quite strong constraints on the Higgs couplings and the heavy scales. Making only some mild assumptions on the high-energy behaviour of the underlying fundamental theory, we were able to show that the precision electroweak data require the Higgs-like scalar to have a $hWW$ coupling very close to the SM one, while the mass of the vector and axial-vector resonances should be quite degenerate and above 4~TeV. Our findings were much more restrictive that the LHC data available at that time. 

The much larger statistics collected in recent years \cite{ATLAS:2022vkf,CMS:2022uhn} has made possible to obtain a more precise experimental measurement of the $hWW$ coupling (in SM units): $\kappa_W=1.023\pm 0.026$~\cite{PDG}.
We can profit this additional information to update our analysis and investigate the numerical impact of the different approximations that were made in Ref.~\cite{ST}.
Thus, we will no-longer consider $\kappa_W$ as a free parameter and will take instead its experimental value as an input. We can then extend the analysis to a broader set of possible interactions. 
While Ref.~\cite{ST} only studied the  
bosonic P-even sector, we will now take into account P-even and P-odd operators.

The oblique parameters $S$ and $T$ can be conveniently computed through two convergent dispersive representations in terms of corresponding spectral functions, which at leading order (LO) are generated by the exchange of massive vector and axial-vector states ($T=0$ at LO). The next-to-leading-order (NLO) contributions are dominated by the lightest two-particle cuts. Corrections from multi-particle cuts involving heavy resonances are kinematically suppressed and have been estimated to be very small \cite{S_Higgsless}. 
For this reason, one can safely disregard any contributions from intermediate fermionic resonances.
Ref.~\cite{ST} computed the leading NLO contributions from two-Goldstone ($\varphi\varphi$) and Higgs-Goldstone ($h\varphi$) intermediate states. In addition, we will also consider the corrections induced by the (light) fermion-antifermion ($\psi \bar{\psi}$) cut.

Our resonance Lagrangian is presented in Section~\ref{sec:lagrangian}, whereas the $S$ and $T$ parameters and their calculation in terms of dispersive representations are discussed in Section~\ref{oblique_parameters}. The LO calculation of the oblique parameters and its phenomenological consequences are shown in Section~\ref{S_at_LO}. Section~\ref{S_at_NLO} contains the NLO calculation of these observables and an extensive discussion of the assumed short-distance constraints, which play a very important role in the theoretical analysis. The phenomenological implications of these estimations are studied in Section~\ref{sec:phenomenology}. Figures~\ref{fig:NLO_2WSR} and \ref{fig:NLO_1WSR} constitute the main result of this work. Finally, we discuss the main  
conclusions in Section~\ref{sec:conclusions}. 
Some technical details of the calculation are explained in Appendices A and B.

\section{The Lagrangian} \label{sec:lagrangian}

Although the expansion in powers of generalized momenta~\cite{Pich:2018ltt,lagrangian,lagrangian_color,Weinberg} is not directly applicable to the resonance theory, one can construct the effective Lagrangian in a consistent phenomenological way, {\it \`a la} Weinberg~\cite{Weinberg}, which interpolates between the low-energy and the high-energy regimes: the appropriate low-energy predictions are generated and a given short-distance behavior is imposed~\cite{Ecker:1988te,Ecker:1989yg}. Then, 
\begin{equation}
\mathcal{L}_{\mathrm{RT}} \,=\, \sum_{\hat d\ge 2}\, \mathcal{L}_{\mathrm{RT}}^{(\hat d)}\,, \label{RT-Lagrangian0}
\end{equation}
where the operators are not ordered according to their canonical dimensions and one must use instead the chiral dimension $\hat d$, which reflects their infrared behavior at low momenta~\cite{Weinberg}. 
Taking into account that here we are interested in the NLO resonance contributions to the $S$ and $T$ parameters from only SM cuts, we only need to consider $\cO(p^2)$ operators with up to one spin-1 bosonic resonance field~\cite{lagrangian,lagrangian_color}. 

Following the notation of Refs.~\cite{lagrangian,lagrangian_color}, the dimension of the resonance representation is indicated with upper and lower indices in the scheme $R_{SU(2)}^{SU(3)}$, where $R$ stands for any of the four possible $J^{PC}$ bosonic states with quantum numbers $0^{++}$ (S), $0^{-+}$ (P), $1^{--}$ (V) and $1^{++}$ (A). The normalization used for the $SU(2)$ triplet resonances is
\begin{equation}
R^{n}_3 \,=\, \frac{1}{\sqrt{2}}\,\sum_{i=1}^3\,\sigma_{i}\,R^n_{3,i}\,, 
\end{equation}
with $\bra \sigma_i\sigma_j\ket_2 = 2\delta_{ij}$ and $\langle\cdots\rangle_2$ indicating  an $SU(2)$ trace.  
In addition, the resonances are classified in Ref.~\cite{lagrangian_color} according to their $SU(3)_C$ color quantum numbers, e.g., for color octets one has  
$R_m^8 \,=\,  \sum_{a=1}^8\,T^a\, R_m^{8,a}$, with $\bra T^a T^b\ket_3 = \delta^{ab}/2$ and $\langle\cdots\rangle_3$ indicating an $SU(3)$ trace. However, colored resonances are irrelevant for the present work: only color-singlet resonances $R_m^1$ contribute to our determination of the oblique parameters. 
To simplify the notation, we 
will denote the $SU(2)$ triplet $R^1_3$ resonance masses as $M_R$. The $SU(2)$ singlet resonances $R^1_1$ will be essentially irrelevant for this work and their masses will be denoted as $M_{R_1}$ when they are later discussed.

The scalar and pseudoscalar resonances do not contribute either to the oblique parameters at the level of accuracy we are considering. They do not generate any tree-level (LO) contribution to $S$ and $T$, while their first NLO contributions originate from cuts involving their heavy masses, which are kinematically suppressed in comparison to the contributions from light-particle cuts that we are going to include in our calculation.

The spin-1 resonances $V$ and $A$ can be described with either a four-vector Proca field $\hat{R}^\mu$ or with an antisymmetric tensor $R^{\mu \nu}$. Both descriptions are equivalent, as they generate the same physical predictions, once proper short-distance constraints are implemented \cite{Ecker:1989yg}. However, they involve different
sets of Goldstone operators without resonance fields that compensate their different scaling with momenta. For simplicity, we keep here both formalisms because, as it was demonstrated in Ref.~\cite{lagrangian}, the sum of tree-level resonance-exchange contributions from the $\cO(p^2)$ resonance Lagrangian with Proca and antisymmetric spin-1 resonances gives the complete (non-redundant and correct) set of predictions for the $\cO(p^4)$ LECs of the electroweak effective theory, without any additional contributions from local operators without resonance fields.

The relevant terms of the electroweak resonance Lagrangian contributing to our NLO calculation of the oblique parameters are~\cite{lagrangian,lagrangian_color}: 
\begin{widetext}
\begin{align}
\Delta \mathcal{L}_{\mathrm{RT}} &=\,\quad 
\sum_{\xi} \left[ i\,\bar\xi \gamma^\mu d_\mu \xi  
 - v \left(\bar{\xi}_L \mY \xi_R + \mbox{h.c.}\right)  
\right]  \,+\, \frac{v^2}{4}\,\left( 1 +\Frac{2\kappa_W}{v} h \right) \bra u_\mu u^\mu\ket_{2} \nonumber \\
&+\,\bra V^1_{3\,\mu\nu} \left( \Frac{F_V}{2\sqrt{2}}  f_+^{\mu\nu} + \Frac{i G_V}{2\sqrt{2}} [u^\mu, u^\nu]  + \Frac{\widetilde{F}_V }{2\sqrt{2}} f_-^{\mu\nu}  +  \Frac{ \widetilde{\lambda}_1^{hV} }{\sqrt{2}}\left[  (\partial^\mu h) u^\nu-(\partial^\nu h) u^\mu \right]   + C_{0}^{V^1_3} J_T^{\mu\nu}  \right) \ket_2 \nonumber \\
&+ \,\bra A^1_{3\,\mu\nu} \left(\Frac{F_A}{2\sqrt{2}}  f_-^{\mu\nu}  + \Frac{ \lambda_1^{hA} }{\sqrt{2}} \left[ (\partial^\mu h) u^\nu-(\partial^\nu h) u^\mu \right] +  \Frac{\widetilde{F}_A}{2\sqrt{2}} f_+^{\mu\nu} +  \Frac{i \widetilde{G}_A}{2\sqrt{2}} [u^{\mu}, u^{\nu}]   +  \widetilde{C}_{0}^{A^1_3}  J_{T}^{\mu\nu} \right) \ket_2 \nonumber \\
&+\,\hat V^1_{1\,\mu} \!\left(\! \widetilde{c}_{\mathcal{T}}  \bra u^\mu \mathcal{T} \ket_2 + \Frac{c^{{\hat{V}}^1_1}}{\sqrt{2}}  \bra J^\mu_V\ket_2  + \Frac{   \widetilde c^{{\hat{V}}^1_1}  }{\sqrt{2}}
\bra J^\mu_A\ket_2  \!\right)\!  
+\, \hat A^1_{1\,\mu} \! \left(\! c_{\mathcal{T}}  \bra u^\mu \mathcal{T} \ket_2 + \Frac{  c^{{\hat{A}}^1_1}  }{\sqrt{2}}  \bra J^\mu_A\ket_2  + \Frac{  \widetilde c^{{\hat{A}}^1_1}  }{\sqrt{2}}  \bra J^\mu_V\ket_2 \!\right)\!.  
\label{eq:Lagr}
\end{align} 
\end{widetext}
The first line shows the non-resonant interactions; the second and third lines contain the vector and axial-vector contributions with an antisymmetric formalism; and the last line displays additional vector and axial-vector operators with the Proca formalism. 
Couplings with (without) a tilde indicate P-odd (P-even) operators.

The Goldstone fields are parametrized through the SU(2) matrix $U = u^2 = \exp{(i \vec{\sigma}\vec{\varphi}}/v)$, where $v= (\sqrt{2} G_F)^{-1/2}=246\;\mathrm{GeV}$ is the EWSB scale,  $u_\mu = -i u^\dagger D_\mu U u^\dagger$ with $D_\mu$ the appropriate covariant derivative,
and $f_\pm^{\mu\nu}$ contain the gauge-boson field strengths.
The known fermions $\psi_i = (u_i, d_i)^T, \, (\nu_i,e_i)^T$ are incorporated through the fields $\xi
= u^\dagger \psi_L + u\psi_R$ and the fermion bilinears $J_{V}^\mu =\bar\xi\gamma^\mu\xi$,
$J_{A}^\mu =\bar\xi\gamma^\mu\gamma_5\xi$ and $J_{T}^{\mu\nu}=\bar\xi\sigma^{\mu\nu}\xi$. The couplings $\widetilde{c}_T$ and $c_T$ in the last line account for the explicit breaking of custodial symmetry, induced by quantum loops with internal $U(1)_Y$ gauge-boson lines, through the SU(2) spurion $\mathcal{T}$. All technical details can be found in Refs.~\cite{lagrangian,lagrangian_color}.

\section{Oblique parameters} \label{oblique_parameters}

From now on we follow the notation of Refs.~\cite{ST,S_Higgsless}. The computation is performed in the Landau gauge, so that the gauge boson propagators are transverse and their self-energies,
\begin{align}
\mathcal{L}_{\mathrm{v.p.}}=& - \frac{1}{2} W^3_\mu\, \Pi^{\mu\nu}_{33}(s)W^3_\nu -\frac{1}{2}B_\mu\,\Pi^{\mu\nu}_{00}(s) B_\nu \nonumber \\ &\quad  - W^3_\mu\, \Pi^{\mu\nu}_{30}(s) B_\nu - W^+_\mu\,\Pi^{\mu\nu}_{WW}(s)W^-_\nu\,, \phantom{\frac{1}{2}}
\end{align}
can be decomposed as
\begin{equation}
\Pi^{\mu\nu}_{ij} (q^2) \,=\, \left( -g^{\mu\nu} +
\frac{q^\mu q^\nu}{q^2}\right)\; \Pi_{ij}(q^2).
\end{equation}
The precise definitions of the $S$ and $T$ {oblique parameters} involve the quantities
\begin{equation}
e_3\,=\, \Frac{g}{g'}  \; \widetilde{\Pi}_{30}(0) \, ,
\qquad
e_1\,=\,
\frac{ \Pi_{33}(0) - \Pi_{WW} (0)}{M_W^2}\,,
\end{equation}
where the tree-level Goldstone contribution has been removed from $\Pi_{30}(s)$ in the form~\cite{Peskin_Takeuchi}:
\begin{equation}
\Pi_{30}(s)\,=\,s\, \widetilde\Pi_{30}(s)\,+\,\frac{g^2 \tan{\theta_W}}{4}\, v^2 \,  .
\end{equation}
The $S$ and $T$ precision observables 
parametrize 
the deviations of $e_3$ and $e_1$ with respect to the SM contributions $e_3^{\rm SM}$ and $e_1^{\rm SM}$, respectively:
\begin{equation}
S\,=\,  \Frac{16\pi}{g^2}\;\big(e_3 - e_3^{\rm SM}\big)\, ,
\quad 
T\,=\, \Frac{4\pi }{
g'^{\, 2} \cos^2{\theta_W} 
}\; \big(e_1-e_1^{\rm SM}\big)  \,.
\end{equation}

As experimental values for $S$ and $T$ we consider the values given by the Particle Data Group~\cite{PDG}, $S=-0.05\pm 0.07$ and $T=0.00 \pm 0.06$ (with a correlation of $0.93$). Note that we are taking the set of values assuming that the oblique parameter $U=0$, which is expected to be suppressed compared to $S$ and $T$.

\subsection{Dispersive representation for $S$ and $T$}\label{dispersive}

A useful dispersive representation for the $S$ parameter was  introduced by Peskin and Takeuchi~\cite{Peskin_Takeuchi}:
\begin{equation}
S \,=\, \Frac{16\pi}{g^2 \tan\theta_W}\; \Int_0^\infty \Frac{{\rm ds}}{s}\; [\rho_S(s)\, -\,\rho_S(s)^{\rm SM} ]\, , \label{Peskin-Takeuchi}
\end{equation}
with the spectral function
\begin{equation}
\rho_S(s) \,=\,\Frac{1}{\pi}\,\mbox{Im}\widetilde{\Pi}_{30}(s)\, .
\end{equation}
The SM one-loop spectral function reads (at lowest order in $g$ and $g'$)
\begin{align}
\rho_S(s)^{\rm SM} =  \frac{g^2 \tan\theta_W}{192\pi^2} \left[ \theta(s)  \!-\!  \left(1\!-\!\frac{m_h^2}{s} \right)^3  \theta \left(s\!-\!m_h^2 \right)  
\hskip 1.2cm\mbox{}
\right. \nonumber \\ 
 \left. +  \sum_{\psi} \theta\!\left(s\!-\! 4m_\psi^2 \right) \sqrt{1\!-\!\frac{4m_\psi^2}{s} }  \left( \frac{6m_\psi^2}{s} 
 \!+\! 
  8 T_3^\psi
 x_\psi \left( 1 \!+\!\frac{2m_\psi^2}{s} 
 \right) \!\right)\! \right] \! , \label{rho_SM}
\end{align}
where there is a sum over all the SM fermion $\psi\overline{\psi}$ cuts and $x_\psi=\frac{1}{2}({\rm B-L})_\psi$ is the corresponding $U(1)_X$ charge of the fermion $\psi$ ($1/6$ for quarks and $-1/2$ for leptons).  
In the SM, and neglecting corrections of $\mathcal{O}(m_h^2/s)$ and $\mathcal{O}(m_\psi^2/s)$, 
the $h\varphi$ and $\varphi\varphi$ contributions to the spectral function $\rho_S(s)$ cancel each other, and a similar cancellation occurs when the $\rho_S$ cuts with up and down components are summed up, for a given fermion doublet
(e.g., $t$ with $b$ for quarks, and $\nu_e$ with $e^-$ for leptons).

For the computation of $T$, we will use the Ward-Takahashi identity worked out in Ref.~\cite{Barbieri:1992dq}. In the Landau gauge, instead of studying the more cumbersome correlators $\Pi_{33}$ and $\Pi_{WW}$, one simply needs to compute the self-energies of the electroweak Goldstones~\cite{Barbieri:1992dq}:
\begin{equation}
e_1 \,=\, \Frac{Z^{(+)}}{Z^{(0)}} \, -\, 1\, \,\, \simeq  \,\,\, \Sigma'^{(0)}(0)\, -\, \Sigma'^{(+)}(0)\, .
\label{eq.Ward-id}
\end{equation}
The constants  $Z^{(+)}$ and $Z^{(0)}$ are the wave-function renormalizations for the charged and neutral Goldstones, respectively. More precisely, they are provided by the derivative of the Goldstone self-energies at zero momentum: $Z^{(i)}=1-\Sigma'^{(i)}(0)$, with $\Sigma'(s)\equiv\mathrm{d}\Sigma(s)/\mathrm{d}s$. This leads to the second identity in~(\ref{eq.Ward-id}), which holds as far as the calculation remains at the NLO.

In Ref.~\cite{ST} we showed that, once proper short-distance conditions have been imposed, the spectral function of the Goldstone self-energy difference,
\begin{equation}
\rho_T(s)\,=\, \frac{1}{\pi}\mbox{Im}[\Sigma^{(0)}(s)-\Sigma^{(+)}(s)]\, , \label{calculation_T_1}
\end{equation}
vanishes at high energies. Hence, one is allowed to recover the low-energy value of the self-energy difference and the $T$ parameter by means of a convergent dispersion relation:
\begin{equation}
T \,=\, \Frac{4\pi}{ 
 g'^{\, 2}  \cos^2{\theta_W} 
} \; \Int_0^\infty \Frac{{\rm ds}}{s^2} \; [\rho_T(s)\, -\, \rho_T(s)^{\rm SM}]\, ,  
\label{eq:T-disp-rel}
\end{equation}
where the SM one-loop spectral function reads (at lowest order in $g$ and $g'$)
\begin{eqnarray}
\rho_T(s)^{\rm SM} &\,=\,& \Frac{3g'^{\, 2}s}{64\pi^2}\;  \bigg[-\theta(s)+\left(1-\frac{m_h^4}{s^2} \right)\theta (s-m_h^2)\bigg] 
\nonumber\\
 &&\hspace*{-2.cm}  +\frac{N_C s }{8\pi^2 v^2} \bigg[ \! (m_t^2\!+\!m_b^2)\beta_{tb} \!-\!m_t^2\beta_{tt}\!-\!m_b^2\beta_{bb}     \!-\! \frac{(m_t^2\!-\!m_b^2)^2}{s}\beta_{tb}\!\bigg]\!, 
\nonumber\\ 
\label{calculation_T_3}
\end{eqnarray}
with $N_C$ the number of colors of the fermion doublet, \ $\beta_{ij}\equiv s^{-1} \lambda^{1/2}(s,m_i^2,m_j^2)\, \theta\left(s-[m_i+m_j]^2\right)$  \ and \  the K\"all\'en function  \ $\lambda(x,y,z)=x^2+y^2+z^2-2 xy -2 yz -2 zx$. 
The SM $h\varphi$ and $\varphi\varphi$ contributions to the spectral function $\rho_T(s)$ cancel each other at high energies. 
A similar high-energy cancellation occurs between the top and bottom components of the $SU(2)$ doublet
because this fermion-loop contribution should vanish in the limit $m_t=m_b$ where custodial symmetry is recovered.
Therefore, $\rho_T(s)^{\mathrm{SM}}$ 
behaves like $\sim 1/s$ at high energies, both for boson and fermion contributions, separately. 
However, the SM fermion cuts generate identical contributions in the BSM extension, except for additional terms involving the custodial-breaking couplings
$c_{\mathcal{T}}$ and $\widetilde{c}_{\mathcal{T}}$. Thus, the fermionic contributions to $T$ are suppressed by additional powers of $g'$, which we will neglect in this work, so $T$ will be solely determined by the bosonic loops.

\section{LO calculation} \label{S_at_LO}

$T$ vanishes at LO ($T_{\mathrm{LO}}=0$), whereas there is a LO contribution to $S$ ($S_{\mathrm{LO}}\ne 0$) given by\footnote{
The tree-level contributions from the Proca operators in 
(\ref{eq:Lagr}) 
can be considered subleading and will be taken into account in the NLO analysis of the next section.}
\begin{equation}
\left. \Pi_{30}(s) \right|_{\mathrm{LO}} =\frac{g^2  \tan{\theta_W} }{4} s  \left(\!\frac{v^2}{s}\!+\!  \frac{F_V^2\!-\!\widetilde{F}_V^2}{M_V^2\!-\!s}  \!-\! \frac{F_A^2\!-\!\widetilde{F}_A^2}{M_A^2\!-\!s} \right) , \label{PI_LO}
\end{equation}
Therefore,
\begin{equation}
S_{\mathrm{LO}} \,=\, 4\pi  \left( \frac{F_V^2-\widetilde{F}_V^2}{M_V^2}  - \frac{F_A^2-\widetilde{F}_A^2}{M_A^2} \right)\, . \label{S_LO_0}
\end{equation}

\subsection{Weinberg Sum Rules}\label{sec:WSRS}

The $W^3B$ correlator $\Pi_{30}(s)$ can be written in terms of  the vector ($R+L$) and axial-vector ($R-L$) two-point functions as~\cite{Peskin_Takeuchi},
\begin{equation}
\Pi_{30}(s)\,=\, \frac{g^2  \tan{\theta_W} }{4} s \left[ \Pi_{VV}(s) - \Pi_{AA}(s) \right] . \label{pi_correlator}
\end{equation}
The assumed chiral symmetry of the underlying electroweak theory implies that this correlator is an order parameter of the EWSB. In asymptotically-free gauge theories it vanishes at short distances as $1/s^3$~\cite{Bernard:1975cd}, implying two superconvergent sum rules, the so-called first and second Weinberg Sum Rules (WSRs)~\cite{WSR}:
\begin{enumerate}
\item $1^{\text{st}}$ WSR (vanishing of the $1/s$ term of $\Pi_{VV}(s) - \Pi_{AA}(s)$). At LO, and from (\ref{PI_LO}) and (\ref{pi_correlator}), one gets~\cite{PRD2}:
\begin{equation}
\left( F_V^2 -\widetilde{F}_V^2  \right) - \left( F_A^2 -\widetilde{F}_A^2 \right) \,=\, v^2\,. \label{1stWSR_LO}
\end{equation}
\item $2^{\text{nd}}$ WSR (vanishing of the $1/s^2$ term of $\Pi_{VV}(s) - \Pi_{AA}(s)$). At LO, and from (\ref{PI_LO}) and (\ref{pi_correlator}), one gets~\cite{PRD2}:
\begin{equation}
\left( F_V^2 -\widetilde{F}_V^2  \right) M_V^2 - \left( F_A^2 -\widetilde{F}_A^2 \right) M_A^2\,=\, 0\,. \label{2ndWSR_LO}
\end{equation}
\end{enumerate}
While the $1^{\text{st}}$ WSR is expected to be also fulfilled in gauge theories with nontrivial ultraviolet (UV) fixed points, the validity of the $2^{\text{nd}}$ WSR depends on the particular type of UV theory considered~\cite{1stWSR}. 

When both WSRs are satisfied, they imply~\cite{PRD2}:
\be\label{eq:FVFA-WSRs}
F_V^2 - \widetilde{F}_V^2\, =\, \frac{v^2 M_A^2}{M_A^2-M_V^2}\, ,
\quad
F_A^2 - \widetilde{F}_A^2 \,=\, \frac{v^2 M_V^2}{M_A^2-M_V^2}\, . 
\ee
Therefore, the differences $F_V^2 - \widetilde{F}_V^2$, $F_A^2 - \widetilde{F}_A^2$ and $M_A^2-M_V^2$ must have the same sign.
In the absence of P-odd couplings, these relations fix $F_V$ and $F_A$ in terms of the vector and axial-vector masses and, moreover, require that $M_A>M_V$. This mass hierarchy remains valid if $\widetilde{F}_V^2 < F_V^2$ and $\widetilde{F}_A^2 < F_A^2$, which is a reasonable working assumption that we will adopt. 
We will also assume that the inequality $M_A>M_V$ is fulfilled in all dynamical scenarios, even when the $2^{\text{nd}}$ WSR does not apply.

\subsection{Phenomenology at LO}

We want to analyze now the implications of the short-distance constraints of (\ref{1stWSR_LO}) and (\ref{2ndWSR_LO}) in the LO prediction of $S$, given in (\ref{S_LO_0}).

If one considers both WSRs, the combinations of resonance couplings $F_V^2 - \widetilde{F}_V^2$ and $F_A^2 - \widetilde{F}_A^2$ are determined by (\ref{eq:FVFA-WSRs}), so one gets
\begin{equation}
S_{\mathrm{LO}} = 4\pi v^2 \!\left(\! \frac{1}{M_V^2} \!+\! \frac{1}{M_A^2} \!\right) = \frac{4\pi v^2}{M_V^2} \!\left(\! 1 \!+\! \frac{M_V^2}{M_A^2} \!\right) . \label{S_LO}
\end{equation}
Therefore, and assuming $M_A > M_V$, the prediction for $S_{\mathrm{LO}}$ is bounded by
\begin{equation}
\frac{4\pi v^2}{M_V^2} \,< \, S_{\mathrm{LO}} \,< \,  \frac{8\pi v^2}{M_V^2}\,. 
\end{equation}

If one considers only the  $1^{\mathrm{st}}$ WSR, and assuming $M_A > M_V$ and $\widetilde{F}_A^{2} < F_A^{2}$, Eq.~(\ref{1stWSR_LO}) allows us to get a lower bound for $S_{\mathrm{LO}}$:
\begin{equation} \label{S_LO_1WSR}
S_{\mathrm{LO}}  = 4\pi \!\left\{ \!\frac{v^2}{M_V^{2}}  \! +\! \left(\!F_A^{2} \!-\! \widetilde{F}_A^{2}\! \right)\!\left( \!\frac{1}{M_V^{2}} \!-\! \frac{1}{M_A^{2}} \!\right) \!\right\}  >  \frac{4\pi v^2}{M_V^{2}} .
\end{equation}

\begin{figure}
\begin{center}
\includegraphics[scale=0.5]{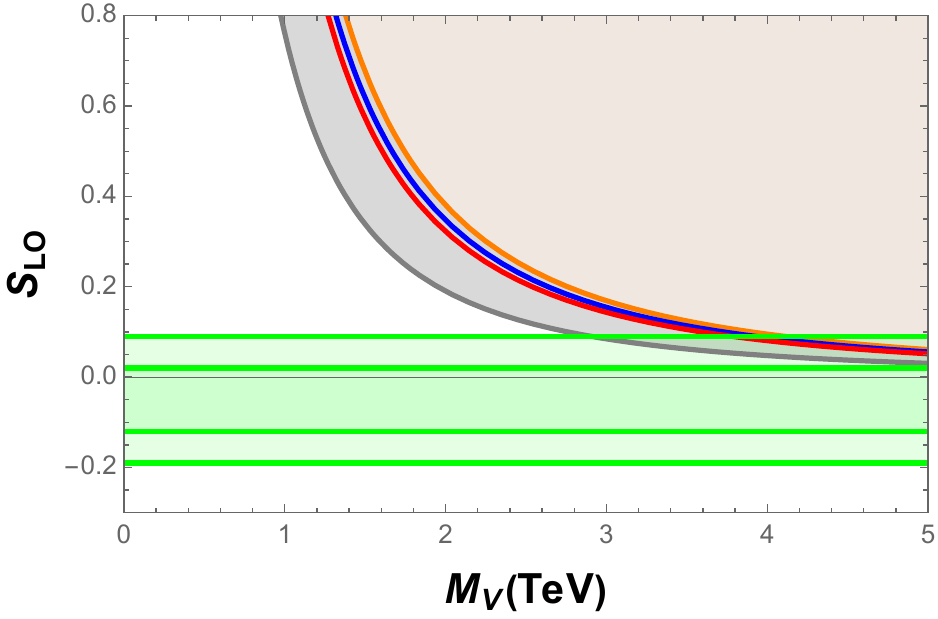}
\caption{{\small
LO predictions for $S$. The green area covers the experimentally allowed region, at 68\% and 95\% CL. The gray region assumes the two WSRs and we indicate explicitly the corresponding lines for $M_A=M_V$ (orange), $M_A=1.1\, M_V$ (blue), $M_A=1.2\, M_V$ (red) and $M_A \to \infty$ (dark gray).  If only the  $1^{\text{st}}$ WSR is considered, the allowed region is given by both, the gray and the brown areas (assuming $M_A > M_V$).
}}
\label{plotSLO}
\end{center}
\end{figure}

These results are identical to the ones we got in Ref.~\cite{ST}. The inclusion of P-odd operators has not changed the LO predictions because the couplings $\widetilde{F}_{V,A}$ get reabsorbed through the relations \eqn{1stWSR_LO} -- \eqn{eq:FVFA-WSRs}.
In Figure~\ref{plotSLO} we show these predictions, together with the experimentally allowed region at 68\% and 95\% CL~\cite{PDG}. The gray area assumes both WSRs and $M_A>M_V$. The colored curves indicate explicitly the predicted results for $M_A=M_V$ (orange), $M_A=1.1\, M_V$ (blue), $M_A=1.2\, M_V$ (red) and $M_A \to \infty$ (dark gray). When only the $1^{\text{st}}$ WSR is considered (and assuming $M_A>M_V$ and $\widetilde{F}_{V,A}^{2} < F_{V,A}^{2}$), the allowed range gets enlarged to the brown region.
Note that the experimental data imply $M_V \!\gsim\! 2.8\,$TeV (95\% CL).\footnote{This procedure is equivalent to the comparison of the theoretical prediction of the electroweak effective theory LEC $\mathcal{F}_1$ with its experimentally allowed region done in Ref.~\cite{PRD2} (top-left plot in Figure~1 of~\cite{PRD2}); consequently, 
this limit updates the bound for $M_V$ that was found there.}

\section{NLO calculation} \label{S_at_NLO}

Taking into account the previous results, it is straightforward to write $\Pi_{30}(s)|_{\mathrm{NLO}}$ as:
\begin{align}
\left. \Pi_{30}(s) \right|_{\mathrm{NLO}}  = &\,\frac{g^2\tan{\theta_W} }{4} s    \left(\frac{v^{2}}{s} \!+\!  \frac{F_{V}^{r\,2}\!-\!\widetilde{F}_V^{r\,2}}{M_{V}^{r\,2}\!-\!s} \right. \nn \\ &\qquad \qquad \phantom{\frac{1}{2}}  \!-\! \frac{F_{A}^{r\,2}\!-\!\widetilde{F}_A^{r\,2}}{M_{A}^{r\,2}\!-\!s} 
+ \overline{\Pi}(s)\Bigg) ,
\label{eq:PiNLO}
\end{align}
where the one-loop contribution from the two-particle cuts is contained in $\overline{\Pi}(s)$. One gets now
\begin{equation}
\widetilde\Pi_{30}(s)|_{\mathrm{NLO}} \!=\! \frac{g^2  \tan{\theta_W} }{4}\!  \left( \! \frac{F_{V}^{r\,2}\!-\!\widetilde{F}_V^{r\,2}}{M_{V}^{r\,2}\!-\!s}  \!-\! \frac{F_{A}^{r\,2}\!-\!\widetilde{F}_A^{r\,2}}{M_{A}^{r\,2}\!-\!s}  \!+\! \overline{\Pi}(s) \!\right) \!,
\end{equation}
and then finally
\begin{equation}
S_{\mathrm{NLO}} = 4\pi \left( \frac{F_{V}^{r\,2}\!-\!\widetilde{F}_V^{r\,2}}{M_{V}^{r\,2}}  \!-\! \frac{F_{A}^{r\,2}\!-\!\widetilde{F}_A^{r\,2}}{M_{A}^{r\,2}}  \right)   + \overline{S}  , \label{S_NLO}
\end{equation}
being $\overline{S}=4\pi \left[ \overline{\Pi}(0)-\overline{\Pi}(0)^{\rm SM}\right]$.

In order to be able to estimate the renormalized couplings of (\ref{S_NLO}), one can consider the high-energy expansion of the contribution from the two-particle cuts,
\begin{align}
\overline{\Pi}(s) &=
\Frac{ v^2}{s} \; \bigg[\delta_{_{\rm NLO}}^{(1)} + \widetilde\delta_{_{\rm NLO}}^{(1)} 
\log{\left(\Frac{-s}{M_V^2}\right)} 
\bigg]  \nn\\ & 
 + \Frac{v^2 \, M_V^2}{s^2}\; \bigg[\delta_{_{\rm NLO}}^{(2)}  + \widetilde\delta_{_{\rm NLO}}^{(2)} 
 \log{\left(\Frac{-s}{M_V^2}\right)} \bigg]  +\cO\bigg(\Frac{1}{s^3}\bigg)\, ,  \label{Pi_expansion}
\end{align}
so that the $1^{\text{st}}$ and $2^{\text{nd}}$ WSRs get modified from their LO expressions in (\ref{1stWSR_LO}) and (\ref{2ndWSR_LO}) to, respectively,
\begin{equation}
 \left( F_V^{r\,2} \!-\! \widetilde{F}_V^{r\,2} \right)  \!-\! \left( F_A^{r\,2}\!-\! \widetilde{F}_A^{r\,2}\right)   = v^2  \left( 1\!+\! \delta_{_{\rm NLO}}^{(1)} \right) \, ,  \label{constraint_1WR}
\end{equation}
\vspace{-0.75cm}
\begin{equation}
 \left( F_V^{r\,2} \!-\! \widetilde{F}_V^{r\,2} \right) M_V^{r\,2} \!-\! \left( F_A^{r\,2}\!-\! \widetilde{F}_A^{r\,2}\right) M_A^{r\,2} = v^2 \,M_V^{r\,2}\,  \delta_{_{\rm NLO}}^{(2)}  \,, \label{constraint_2WR}
\end{equation}
plus the additional conditions 
$\widetilde\delta_{_{\rm NLO}}^{(1)}=0$ ($1^{\text{st}}$ WSR) and $\widetilde\delta_{_{\rm NLO}}^{(2)}=0$ ($2^{\text{nd}}$ WSR).
Note that in (\ref{Pi_expansion}) we have assumed 
that $\overline{\Pi}(s)$ vanishes 
at high energies, since, as we will see, well-behaved form factors are considered.

If we assume the validity of both WSRs, it is then possible to determine the combination of NLO resonance couplings appearing in (\ref{S_NLO}), since from (\ref{constraint_1WR}) and (\ref{constraint_2WR}) one can get:
\begin{align}
F_{V}^{r\,2} \!-\! \widetilde{F}_V^{r\,2} &=  \frac{v^2 \,M_{A}^{r\,2}}{M_{A}^{r\,2}\!-\!M_{V}^{r\,2}}\;
        \left(1\!+\!\delta_{_{\rm NLO}}^{(1)}\!-\!\frac{M_{V}^{r\,2}}{M_{A}^{r\,2}}\,\delta_{_{\rm NLO}}^{(2)}  \right)  ,
\nn\\[10pt]
F_A^{r\,2}\!-\! \widetilde{F}_A^{r\,2}&=  \frac{v^2 \,M_{V}^{r\,2}}{M_{A}^{r\,2}\!-\!M_{V}^{r\,2}}\;
        \Bigl(1\!+\!\delta_{_{\rm NLO}}^{(1)}\!-\!\delta_{_{\rm NLO}}^{(2)} \Bigr)  ,
\label{FAr}
\end{align} 
and therefore
\begin{equation}
S_{\mathrm{NLO}}  \!=\!  4\pi v^2  \!\bigg[\!\Frac{1}{M_{V}^{r\,2}} +\Frac{1}{M_{A}^{r\,2}}\!\bigg]\!
  \left(\!1\!+\!\delta_{\rm NLO}^{(1)}
\!-\!  \frac{M_{V}^{r\,2}\, \delta_{\rm NLO}^{(2)}}{M_{V}^{r\,2}\!+\!M_{A}^{r\,2}}\! \right) \! +\!\overline{S} .
\label{eq.1+2WSR}
\end{equation}
Note the similarity between the NLO results of (\ref{FAr}) and (\ref{eq.1+2WSR}) with the LO results of (\ref{eq:FVFA-WSRs}) and (\ref{S_LO}).

If we only consider the  $1^{\text{st}}$ WSR, it is possible to get at least a lower bound 
for $S$:
\begin{eqnarray}
S_{\mathrm{NLO}} &=& 4\pi \!\left\{\!
\frac{v^2}{M_V^{r\, 2}} \! \left( \!1\!+\!\delta_{_{\rm NLO}}^{(1)}\right) \!+\! \left(\!F_A^{r\,2} \!-\! \widetilde{F}_A^{r\,2} \!\right)\!\!\left(\! \frac{1}{M_V^{r\, 2}} \!-\! \frac{1}{M_A^{r\, 2}}\! \right)\!\! \right\}   \nonumber \\ 
&+& \overline{S}\quad > \quad\frac{4\pi v^2}{M_V^{r\, 2}}  \left( 1+\delta_{_{\rm NLO}}^{(1)}\right)  + \overline{S} \, .
\label{eq.NLO-S+1WSR}
\end{eqnarray} 
Note that we have assumed that $M_A^{r\, 2} > M_V^{r\, 2}$ and that $\widetilde{F}_A^{r\,2} < F_A^{r\,2}$. Again, notice the similarity between the NLO result of (\ref{eq.NLO-S+1WSR}) and the LO result of (\ref{S_LO_1WSR}).

\subsection{\boldmath Custodial-breaking corrections to $S$}

The LO results shown in (\ref{PI_LO}) receive small tree-level corrections from the custodial-breaking operators 
with coefficients $c_{\mathcal{T}}$ and $\widetilde{c}_{\mathcal{T}}$, appearing in the Lagrangian 
(\ref{eq:Lagr}). 
The corresponding contributions to $\left.\Pi_{30}(s) \right|_{\mathrm{LO}}$ and $S_{\mathrm{LO}}$ are given by 
\begin{align}
\left.\Delta \Pi_{30}(s) \right|_{\mathrm{LO}} =& -\frac{g^2  \tan{\theta_W} }{4} s  \left[\frac{2v^2}{s} \left(  \sigma_V+\sigma_A \right)  \right. \nn \\ & \qquad \qquad \left. +\frac{v^2 \sigma_V}{M_V^2-s} + \frac{v^2 \sigma_A}{M_A^2-s} \right] \,, \label{PI_LO_tau} \\
\Delta S_{\mathrm{LO}} =&  -4\pi v^2 \left(\frac{\sigma_V}{M_V^2} + \frac{ \sigma_A}{M_A^2} \right) \,, \label{S_LO_tau}
\end{align}
where we have introduced 
the dimensionless combinations of parameters $\sigma_V \equiv g^2  \tan^2{\theta_W}\, \widetilde{c}_{\mathcal{T}}^2/(v^2 M_V^2)$ and $\sigma_A\equiv g^2  \tan^2{\theta_W} \,c_{\mathcal{T}}^2/(v^2 M_A^2)$. Assuming that $c_{\mathcal{T}} \sim \widetilde{c}_{\mathcal{T}} \sim v^2$ and $M_R \sim 1\,$TeV, one gets $\sigma_V \sim \sigma_A \sim 10^{-2}\ll 1$. 
Moreover, note that the new contributions of (\ref{PI_LO_tau}) and (\ref{S_LO_tau}) are subleading in $g'$ compared to (\ref{PI_LO}) and (\ref{S_LO_0}), respectively.\footnote{The couplings $c_{\mathcal{T}}$ and $\widetilde{c}_{\mathcal{T}}$ account for custodial-breaking effects of $\cO(g')$. Therefore, $\sigma_{V,A}$ are of $\cO(g'^{\, 2} )$.} 
That is why these custodial-breaking contributions have not been taken into account in our LO analysis. We will include the small corrections (\ref{PI_LO_tau}) and (\ref{S_LO_tau}) together with the NLO contributions. Terms of $\cO(\widetilde{c}_{\mathcal{T}}^4)$, $\cO(c_{\mathcal{T}}^4)$, $\cO(\widetilde{c}_{\mathcal{T}}^2 \,c_{\mathcal{T}}^2)$ and higher are tiny and will be neglected in the NLO analysis.

The custodial-breaking contributions of \eqn{PI_LO_tau} can be reabsorbed into $v^{2}$, $F_V^{r\,2} - \widetilde{F}_V^{r\,2}$ and $F_A^{r\,2} - \widetilde{F}_A^{r\,2}$, in \eqn{eq:PiNLO}, via the redefinitions: 
\begin{align}
v^{2}  \quad\rightarrow\quad& v^{2} -2v^2 \left(  \sigma_V+\sigma_A \right) \,, \nonumber \\
F_V^{r\,2} - \widetilde{F}_V^{r\,2}  \quad\rightarrow\quad&  F_V^{r\,2} - \widetilde{F}_V^{r\,2} - v^2 \sigma_V \,, \nonumber \\
 F_A^{r\,2} - \widetilde{F}_A^{r\,2}  \quad\rightarrow\quad & F_A^{r\,2} - \widetilde{F}_A^{r\,2} + v^2 \sigma_A \,.
\end{align}
Therefore, the custodial-breaking corrections are already included in the NLO expressions (\ref{eq.1+2WSR}) and (\ref{eq.NLO-S+1WSR}).

\subsection{Bosonic cuts}
\subsubsection{$\varphi\varphi$ cut}

The two-Goldstone ($\varphi\varphi$) contribution to the spectral function $\rho_S(s)$ can be written in terms of the corresponding vector (VFF) and axial-vector (AFF) form factors,
\begin{equation}
\rho_S(s)|_{\varphi\varphi}  = \theta(s)\; \Frac{g^2\tan\theta_W}{192\pi^2}\;\left(  |\mF^V_{\varphi\varphi}(s)|^2\!-\! |\mF^A_{\varphi\varphi}(s)|^2 \right)  , \label{rho_phiphi}
\end{equation}  
which are defined through
\begin{eqnarray}
\langle \varphi^+ \,\varphi^-|\frac{\delta S_{\mathrm{RT}}}{\delta v^3_\mu }|0\rangle &=& (p_{\varphi^+}-p_{\varphi^-})^\mu \, \mF^V_{\varphi\varphi}(q^2)\, , \nonumber\\
\langle  \varphi^+ \, \varphi^-|\frac{\delta S_{\mathrm{RT}}}{\delta a^3_\mu }|0\rangle &=& (p_{\varphi^+}-p_{\varphi^-})^\mu \,  \mF^A_{\varphi\varphi}(q^2)\, , \end{eqnarray}
with $q=p_{\varphi^+}+p_{\varphi^-}$, where $S_{\mathrm{RT}}$ is the action of the electroweak resonance theory. The sources $v_\mu=v_\mu^a\sigma^a/2$ and $a_\mu=a_\mu^a\sigma^a/2$ incorporate the $U(1)_Y$ and $SU(2)_L$ gauge bosons $B_\mu$ and $W^a_\mu$: $v_\mu+a_\mu= 
-g'B_\mu \sigma^3/2$ and $v_\mu-a_\mu= 
-g W^a_\mu \sigma^a/2$. 
The VFF and AFF were given in Ref.~\cite{lagrangian}:\footnote{Notice the typo in Ref.~\cite{lagrangian}, where the $\varphi\varphi$ AFF carries an additional $(-1)$ global factor.}
\begin{align}
\mF^V_{\varphi\varphi}(s) =&\, 1\, +\,\Frac{F_V\,G_V}{v^2}\,\Frac{s}{M_V^2-s}\,   +\,\Frac{\widetilde{F}_A\,\widetilde{G}_A}{v^2}\,\Frac{s}{M_A^2-s}  \,, \nonumber \\
\mF^A_{\varphi\varphi}(s)=&\, -  \Frac{\widetilde{F}_V\, G_V}{v^2}\,\Frac{s}{M_V^2-s} \, -\, \Frac{F_A\,\widetilde{G}_A}{v^2}\,\Frac{s}{M_A^2-s} \,. \label{FF_phiphi}
\end{align}

\subsubsection{$h \varphi $ cut}

The Higgs-Goldstone ($h \varphi $) contribution to the spectral function can be 
also written in terms of the corresponding vector and axial form factors,
\begin{align}
\rho_S(s)|_{h\varphi}  = &\, \theta(s-m_{h}^2) \,\Frac{g^2\tan\theta_W}{192\pi^2} \left(1-\Frac{m_{h}^2}{s}\!\right)^3  \nonumber \\ & \qquad \quad \phantom{\Bigg(}\times \left( |\mF^V_{h\varphi}(s)|^2- |\mF^A_{h\varphi}(s)|^2 \right) , \label{rho_hphi}
\end{align}
which are  
defined through
\begin{eqnarray}
\langle  \varphi^0 \, h|\frac{\delta S_{\mathrm{RT}}}{\delta v^3_\mu }|0\rangle &=&  i P_T^{\mu\nu}(q)\, (p_{\varphi^0}-p_h)_\nu\, \mF^V_{h\varphi}(q^2)\, ,\nonumber\\
\langle  \varphi^0 \, h|\frac{\delta S_{\mathrm{RT}}}{\delta a^3_\mu }|0\rangle &=& i P_T^{\mu\nu}(q)\, (p_{\varphi^0}-p_h)_\nu\, \mF^A_{h\varphi}(q^2)\, ,  \end{eqnarray} 
with $P_T^{\mu\nu}(q) = g^{\mu\nu} - q^\mu q^\nu/q^2$. They are 
given by:
\begin{align}
\mF^V_{h\varphi}(s) =& \,-\, \kappa_W  \left( \Frac{\widetilde{F}_A\, \lambda_1^{hA}}{\kappa_W \,v}\Frac{s}{M_A^2\!-\!s} \! +\! \Frac{F_V\,\widetilde{\lambda}_1^{hV}}{\kappa_W \,v}\Frac{s}{M_V^2\!-\!s} \right) , \nonumber  \\
\mF^A_{h\varphi}(s)=&\, \kappa_W\left( 1 +\Frac{F_A\,\lambda_1^{hA}}{\kappa_W\, v}\Frac{s}{M_A^2\!-\!s}\!   +\!\Frac{\widetilde{F}_V\,\widetilde{\lambda}_1^{hV}}{\kappa_W \,v}\Frac{s}{M_V^2\!-\!s} \right) . \label{FF_hphi}
\end{align}

\subsubsection{High-energy constraints and contributions to $S$}

As it can be observed in (\ref{FF_phiphi}) and (\ref{FF_hphi}), all the four form factors we have just introduced are non-zero at $s=q^2\to\infty$, implying an unacceptable bad UV behaviour of the spectral function $\rho_S(s)$. Following the same procedure used in Ref.~\cite{ST}, we require these form factors to vanish at very high energies, which enforces the following short-distance conditions~\cite{PRD2}:
\begin{align}
v^2 \!-\! &F_V\,G_V   \!-\!\widetilde{F}_A\,\widetilde{G}_A = 0\, ,  \qquad  
\widetilde{F}_V\, G_V   \!+\! F_A\,\widetilde{G}_A = 0\, , \nonumber \\ 
\widetilde{F}_A\, \lambda_1^{hA}&  \!+\! F_V\,\widetilde{\lambda}_1^{hV} = 0\, ,\qquad  
\kappa_W\,v  \!-\!F_A\,\lambda_1^{hA}   -\widetilde{F}_V\,\widetilde{\lambda}_1^{hV} = 0\, . 
\label{constraints_FF} 
\end{align}
These relations determine the couplings $G_V$, $\widetilde G_A$, $\lambda_1^{hA}$ and $\widetilde{\lambda}_1^{hV}$ in terms of the remaining  
parameters~\cite{QCD23}: 
\begin{align}
\frac{G_V}{F_A} = - \frac{\widetilde G_A}{\widetilde{F}_V} = \frac{\lambda_1^{hA} v }{\kappa_W F_V} = -\frac{\widetilde{\lambda}_1^{hV} v}{\kappa_W \widetilde{F}_A} = \frac{v^2}{F_V F_A - \widetilde{F}_V \widetilde{F}_A}\, . \label{constraints_summary}
\end{align}
Once these determinations are used, the form factors of (\ref{FF_phiphi}) and (\ref{FF_hphi}) can be written as:\footnote{In the absence of P-odd operators, the UV conditions of (\ref{constraints_FF}) imply $\mF^V_{\varphi\varphi}(s)=M_V^2/(M_V^2-s)$ and $\mF^A_{h\varphi}(s)=\kappa_W\,M_A^2/(M_A^2-s)$, while $\mF^A_{\varphi\varphi}(s)= \mF^V_{h\varphi}(s)=0$. Therefore, in Ref.~\cite{ST} it was possible to determine $\rho_S(s)|_{\varphi\varphi + h\varphi}$ in terms of only three parameters ($\kappa_W$, $M_V$ and $M_A$). 
}
\begin{eqnarray}
\mF^V_{\varphi\varphi}(s) & = & \frac{M_V^2}{M_V^2-s} + \Omega\; \phi(s)\, ,
\nonumber\\
\mF^A_{\varphi\varphi}(s) & =&  -\frac{F_A}{\widetilde F_A}\, \Omega\; \phi(s)\, ,
\nonumber\\
\mF^V_{h\varphi}(s) & = & \kappa_W\,\frac{F_V}{\widetilde F_V}\, \Omega\; \phi(s)\, ,
\nonumber\\
\mF^A_{h\varphi}(s) & = & \kappa_W \left\{\frac{M_A^2}{M_A^2-s} - \Omega\; \phi(s)\right\} , \label{FF1}
\end{eqnarray}
where
\begin{equation}
\Omega\,  =\, \frac{\widetilde F_V\widetilde F_A}{F_V F_A - \widetilde F_V\widetilde F_A} \label{FF2}
\end{equation}
and
\begin{equation}
\phi(s)\, =\, \frac{s}{M_V^2\!-\!s} - \frac{s}{M_A^2\!-\!s}\, =\,
\frac{(M_A^2\!-\!M_V^2)\, s}{(M_V^2\!-\!s) (M_A^2\!-\!s)}\, . \label{FF3}
\end{equation}
Therefore,
\begin{align}
&\rho_S(s)|_{\varphi\varphi+h\varphi} \, = \, \theta(s)\, \Frac{g^2\tan\theta_W}{192\pi^2} \times \nonumber \\
&\quad \Bigg\{ 
\left( \frac{M_V^2}{M_V^2-s}\right)^2 -\kappa_W^2 \left( \frac{M_A^2}{M_A^2-s}\right)^2 \nonumber \\
& \quad
+\, 2\,\Omega\;\phi(s)\left[ \frac{M_V^2}{M_V^2-s} + \kappa_W^2\, \frac{M_A^2}{M_A^2-s}\right]
\nonumber\\   
&\quad +\, \Omega^2\;\phi^2(s)\left[ \left(1\!-\!\frac{F_A^2}{\widetilde F_A^2}\right)
- \kappa_W^2 \left(1\!-\!\frac{F_V^2}{\widetilde F_V^2}\right)\right] \Bigg\} , \label{rho_phiphi_hphi}
\end{align}
where we have neglected corrections of $\mathcal{O}\left( m_h^2/s \right)$.

Since the resulting form factors of (\ref{FF1}) fall as $1/s$ at very high energies, $\rho_S(s)|_{\varphi\varphi + h\varphi} \sim 1/s^2$ when $s\to\infty$, see (\ref{rho_phiphi_hphi}), and, consequently,
\begin{equation}
\widetilde\delta_{_{\rm NLO}}^{(1)} \big|_{\varphi\varphi+h\varphi} = 0\, .   
\label{eq:delta1tilde_scalars}
\end{equation}

Although we consider the experimental measurement for $\kappa_W$, the remaining six free parameters in (\ref{rho_phiphi_hphi}) ($M_V$, $F_V$, $\widetilde{F}_V$, $M_A$, $F_A$, $\widetilde{F}_A$) make mandatory the use of approximations before analysing the phenomenology of our results. Consequently, we consider two different approaches.

{\bf Approach A [P-even].} In this first approach we neglect the odd-parity couplings, that is, $\widetilde{F}_V =\widetilde{F}_A = \widetilde{G}_A =\widetilde{\lambda}_1^{hV}=0$, which translates into $\Omega =0$ in (\ref{rho_phiphi_hphi}), so that
\begin{align}
\rho_S(s)|_{\varphi\varphi+h\varphi}^{\mathrm{\bf{A}}} & = \, \theta(s)\, \Frac{g^2\tan\theta_W}{192\pi^2} \times \nonumber \\
& \Bigg\{ 
\left( \frac{M_V^2}{M_V^2-s}\right)^2 -\kappa_W^2 \left( \frac{M_A^2}{M_A^2-s}\right)^2  \Bigg\} , \label{rho_phiphi_hphiA}
\end{align}
where again we have neglected corrections of $\mathcal{O}\left( m_h^2/s \right)$. The corresponding contribution to $\overline{S}$ is given by:
\begin{align}
\overline{S}\big|_{\varphi\varphi+h\varphi}^{\mathrm{\bf{A}}}  \!=& \frac{1}{12\pi}\! \left\{ \!\left(\log \frac{M_V^2}{m_h^2} \!-\!\frac{17}{6}  \right) \!-\! \kappa_W^2 \!\left( \log \frac{M_A^2}{m_h^2}\!-\! \frac{17}{6} \! \right) \!\right\}\,    , \label{S_1WSR}
\end{align}
and the different terms contributing to the high-energy expansion of $\overline{\Pi}(s)$ in (\ref{Pi_expansion}) are:
\begin{align}
\delta_{_{\rm NLO}}^{(1)}\big|_{\varphi\varphi+h\varphi}^{\mathrm{\bf{A}}}=&\,\frac{M_V^2}{48\pi^2v^2} \left( 1 \!-\! \kappa_W^2 \frac{M_A^2}{M_V^2}  \right) \,, \label{delta1_bosonic_approachA} \\
\delta_{_{\rm NLO}}^{(2)}\big|_{\varphi\varphi+h\varphi}^{\mathrm{\bf{A}}}=&\,\frac{M_V^2}{48\pi^2v^2} \!\left(\! 1\!-\!\kappa_W^2\frac{M_A^4}{M_V^4}\! \left[\! 1\!+\! \log \frac{M_A^2}{M_V^2}\right]\!  \right) \!,  \label{delta2_bosonic_approachA} \\
\widetilde\delta_{_{\rm NLO}}^{(2)}\big|_{\varphi\varphi+h\varphi}^{\mathrm{\bf{A}}} =&\,  \frac{M_V^2}{48\pi^2v^2} \left( -1 \!+\!\kappa_W^2 \frac{M_A^4}{M_V^4}  \right) \,. \label{tildedelta2_bosonic_approachA} 
\end{align}
In (\ref{S_1WSR})-(\ref{tildedelta2_bosonic_approachA}) we have neglected corrections of $\mathcal{O}\left( m_h^2/M_V^2 \right)$ and $\mathcal{O}\left( m_h^2/M_A^2 \right)$.

In case of considering both WSRs, these results and (\ref{eq.1+2WSR}) allow us to obtain the contributions from the bosonic cuts in Approach A: 
\begin{align}
&S_{\mathrm{NLO}}  \!=\!  4\pi v^2 \! \bigg(\!\frac{1}{M_{V}^{r\,2}} \!+\!\frac{1}{M_{A}^{r\,2}}\!\bigg) \!+\! \left. S_{\mathrm{NLO}}\right|_{\varphi\varphi, h\varphi}\!+\! \left. S_{\mathrm{NLO}}\right|_{\psi\bar\psi} ,  \label{SNLOA_1} \\
& \left.S_{\mathrm{NLO}}\right|_{\varphi\varphi, h\varphi}^{\mathrm{\bf{A}}}  \!=\! \frac{1}{12\pi} \left[ \left(1-\kappa_W^2\right)  \left( \log \frac{M_V^2}{m_h^2} -\frac{11}{6} \right) \right. \nonumber \\&\qquad \quad  \qquad  \qquad \left. +\kappa_W^2 \left( \frac{M_A^2}{M_V^2}-1 \right) \log \frac{M_A^2}{M_V^2}  \right]\,,  \label{SNLOA}
\end{align}
Note that in (\ref{SNLOA}) the contributions coming from (\ref{delta1_bosonic_approachA}) and (\ref{delta2_bosonic_approachA}) are included and the only additional ingredient would be the vanishing of (\ref{tildedelta2_bosonic_approachA}) enforced by the $2^{\mathrm{nd}}$ WSR.

If only the $1^{\text{st}}$ WSR is imposed, inserting the results of (\ref{S_1WSR}) and (\ref{delta1_bosonic_approachA}) in (\ref{eq.NLO-S+1WSR}) allows us to obtain a lower bound for $S$ in case of assuming only bosonic cuts:
\begin{align}
 & S_{\mathrm{NLO}}   > \frac{4\pi v^2}{M_V^{r\, 2}} + 
 \left.\Delta S_{\mathrm{NLO}}\right|_{\varphi\varphi, h\varphi}+
 \left.\Delta S_{\mathrm{NLO}}\right|_{\psi\bar\psi}
 \,,
 \label{S1WSR_A_1} \\
 &\left.\Delta S_{\mathrm{NLO}}\right|_{\varphi\varphi, h\varphi}^{\mathrm{\bf{A}}} = 
  \frac{1}{12\pi}
\left[     \bigg(1-\kappa_W^2 \bigg) \bigg(\log\frac{M_V^2}{m_{h}^2}-\frac{11}{6}\bigg) \right. \nonumber \\ & \qquad \left.
 - \,\kappa_W^2\, \bigg(\log \frac{M_A^2}{M_{V}^2}-1
 + \frac{M_A^2}{M_V^2}\bigg) \right] \,. \label{S1WSR_A}
\end{align}
Note again that in (\ref{S1WSR_A}) the contribution coming from (\ref{delta1_bosonic_approachA}) is included. All the results of Approach A presented here correspond to the ones reported in Ref.~\cite{ST}, where only P-even operators and bosonic contributions were analyzed.

Be aware of the different definitions
\begin{eqnarray}
\left.S_{\rm NLO}\right|_{\rm cut}&=& \left[\overline{S}+4\pi v^2\left(\frac{1}{M_V^{r\, 2}}+\frac{1}{M_A^{r\, 2}}\right)\delta^{(1)}_{\rm NLO} \right. 
\nonumber\\
&& 
\left.\mbox{} - \frac{4\pi v^2\delta^{(2)}_{\rm NLO}}{M_A^{r\, 2}} \right]_{\rm cut}\, ,
\label{eq:Snlo12}
\end{eqnarray}
in~(\ref{SNLOA_1}) for two WSRs, and
\begin{equation}\label{eq:Snlo1}
\left.\Delta S_{\rm NLO}\right|_{\rm cut}\, =\,\left[\overline{S}+\frac{4\pi v^2 \delta^{(1)}_{\rm NLO}  }{M_V^{r\, 2}} \right]_{\rm cut}\, ,
\end{equation} 
in~(\ref{S1WSR_A_1}) for the 1st WSR case.

{\bf Approach B [P-odd/even].} In this second approach we consider the odd-parity couplings to be subleading. Therefore, in (\ref{rho_phiphi_hphi}) we perform an expansion in $\widetilde{F}_{V,A}/F_{V,A}$, so that 
\begin{equation}
\Omega\, =\, \frac{\widetilde F_V \widetilde F_A}{F_V F_A} + \mathcal{O} \left(\!\frac{\widetilde F_{V,A}^4}{F_{V,A}^4}\!\right)\, , \label{expansion}
\end{equation}
and one finds:
\begin{align}
&\rho_S(s)|_{\varphi\varphi+h\varphi}^{\mathrm{\bf{B}}}  = \rho_S(s)|_{\varphi\varphi+h\varphi}^{\mathrm{\bf{A}}} \,+ \, \theta(s)\, \Frac{g^2\tan\theta_W}{192\pi^2} \times \nonumber \\
 & \quad \frac{s (M_A^2\!-\!M_V^2)}{(M_A^2\!-\!s)^2 (M_V^2\!-\!s)^2}\left\{
 (M_A^2\!-\!M_V^2) s \left[\kappa_W^2 \frac{\widetilde F_A^2}{F_A^2}
\!-\!\frac{\widetilde F_V^2}{F_V^2}\right] \right.
\nonumber \\
& \quad 
 + 2\,\frac{\widetilde F_A \widetilde F_V}{F_A F_V}\,
\left\{ M_A^2\, \left[ (1+\kappa_W^2) M_V^2 - \kappa_W^2 s\right] - M_V^2 s\right\} \Bigg\} 
\nonumber\\ 
&\quad +\! \mathcal{O}\!\left(\!\frac{\widetilde{F}^4_{V,A}}{F^4_{V,A}}\!\right)\, , \label{rho_phiphi_hphiB}
\end{align}
where again we have neglected corrections of $\mathcal{O}\left( m_h^2/s \right)$.

The corresponding contribution to $\overline{S}$ is given by:
\begin{align}
&\overline{S}\big|_{\varphi\varphi+h\varphi}^{\mathrm{\bf{B}}}  \!=\!
\overline{S}\big|_{\varphi\varphi+h\varphi}^{\mathrm{\bf{A}}} \!+\!
\frac{1}{12\pi}\! \left\{ 2\!\left(\!\frac{\widetilde{F}_V}{F_V}\!-\!\frac{\widetilde{F}_A}{F_A}\! \right)\! \left(\!\frac{\widetilde{F}_V}{F_V}\!+\! \kappa_W^2 \frac{\widetilde{F}_A}{F_A}\!\right) \right. \nonumber \\ &
 + \frac{M_V^2}{M_A^2\!-\!M_V^2}\log \frac{M_A^2}{M_V^2}
\left[ \frac{\widetilde{F}_V}{F_V} \left( 2 \frac{\widetilde{F}_A}{F_A} \!-\!\left( 1\!+\! \frac{M_A^2}{M_V^2}\right) \frac{\widetilde{F}_V}{F_V} \right)\right. \nonumber \\ &
\left.\left.  -
\kappa_W^2  \frac{\widetilde{F}_A}{F_A} \left( 2 \frac{M_A^2}{M_V^2} \frac{\widetilde{F}_V}{F_V} \!-\!\left( 1\!+\! \frac{M_A^2}{M_V^2}\right) \frac{\widetilde{F}_A}{F_A}\right) \right]
 \right\} \!+\! \mathcal{O}\!\left(\!\frac{\widetilde{F}^4_{V,A}}{F^4_{V,A}}\!\right) 
\!   , \label{S_1WSR_B}
\end{align}
and the different terms contributing to the high-energy expansion of $\overline{\Pi}(s)$ in (\ref{Pi_expansion}) are:
\begin{widetext}
\begin{align}
&\delta_{_{\rm NLO}}^{(1)}\big|_{\varphi\varphi+h\varphi}^{\mathrm{\bf{B}}}=
\delta_{_{\rm NLO}}^{(1)}\big|_{\varphi\varphi+h\varphi}^{\mathrm{\bf{A}}}+
\,\frac{M_V^2}{48\pi^2v^2}
\left\{ \frac{\widetilde{F}_V}{F_V} \left[ 2 \frac{\widetilde{F}_A}{F_A} - \left( 1 + \frac{M_A^2}{M_V^2}\right) \frac{\widetilde{F}_V}{F_V} \right]  -
\kappa_W^2 \frac{\widetilde{F}_A}{F_A} \left[ 2 \frac{M_A^2}{M_V^2} \frac{\widetilde{F}_V}{F_V} - \left( 1 + \frac{M_A^2}{M_V^2}\right) \frac{\widetilde{F}_A}{F_A} \right] \right. \nn \\ 
&\left. \qquad \qquad  \qquad \qquad  
+ \frac{2M_A^2 }{M_A^2\!-\!M_V^2} \left( \frac{\widetilde{F}_V}{F_V} - \frac{\widetilde{F}_A}{F_A} \right)
\left( \frac{\widetilde{F}_V}{F_V} + \kappa_W^2 \frac{\widetilde{F}_A}{F_A} \right)\log \frac{M_A^2}{M_V^2} \right\}
\!+\! \mathcal{O}\!\left(\!\frac{\widetilde{F}^4_{V,A}}{F^4_{V,A}}\!\right) 
 \,,  \label{delta1_bosonic_approachB} \\
&\delta_{_{\rm NLO}}^{(2)}\big|_{\varphi\varphi+h\varphi}^{\mathrm{\bf{B}}}=
\delta_{_{\rm NLO}}^{(2)}\big|_{\varphi\varphi+h\varphi}^{\mathrm{\bf{A}}}+
\,\frac{M_V^2}{48\pi^2v^2} \left\{
\frac{\widetilde{F}_V}{F_V} \left( 2 \frac{\widetilde{F}_A}{F_A} - \frac{\widetilde{F}_V}{F_V} - \frac{M_A^4}{M_V^4} \frac{\widetilde{F}_V}{F_V}\right) + \kappa_W^2 \frac{\widetilde{F}_A}{F_A} \left[ \frac{\widetilde{F}_A}{F_A} + \frac{M_A^4}{M_V^4} \left( \frac{\widetilde{F}_A}{F_A} - 2 \frac{\widetilde{F}_V}{F_V} \right) \right]
\right. \nn \\
& \left. 
+\frac{M_A^4}{\left(\!M_A^2\!-\!M_V^2 \!\right)\! M_V^2}\!\left[\!
\frac{\widetilde{F}_V}{F_V} \!\left(\! 3 \frac{\widetilde{F}_V}{F_V} \!-\!2 \frac{\widetilde{F}_A}{F_A}\!-\!\frac{M_A^2}{M_V^2} \frac{\widetilde{F}_V}{F_V} \!\right)\! 
\!+\!\kappa_W^2 \frac{\widetilde{F}_A}{F_A}\! \left(\! 4 \frac{\widetilde{F}_V}{F_V} \!-\!3 \frac{\widetilde{F}_A}{F_A} \!+\! \frac{M_A^2}{M_V^2} \!\left(\! \frac{\widetilde{F}_A}{F_A} \!-\! 2\frac{\widetilde{F}_V}{F_V} \!\right)\! \right)\!
\right]\!
\log \!\frac{M_A^2}{M_V^2}\!\right\} \!+\! \mathcal{O}\!\!\left(\!\frac{\widetilde{F}^4_{V,A}}{F^4_{V,A}}\!\right)\! ,
  \label{delta2_bosonic_approachB} \\ 
&\widetilde\delta_{_{\rm NLO}}^{(2)}\big|_{\varphi\varphi+h\varphi}^{\mathrm{\bf{B}}} =\, 
\widetilde\delta_{_{\rm NLO}}^{(2)}\big|_{\varphi\varphi+h\varphi}^{\mathrm{\bf{A}}} + 
\frac{M_A^2 \!-\! M_V^2}{48\pi^2v^2} \left[ \left( \frac{M_A^2}{M_V^2}\!-\!1 \right)  
\left( \frac{\widetilde{F}_V^2}{F_V^2}\!-\!\kappa_W^2  \frac{\widetilde{F}_A^2}{F_A^2} \right) \!+\!  \frac{2\widetilde{F}_V \widetilde{F}_A}{F_V F_A} \left( 1 \!+\! \kappa_W^2 \frac{M_A^2}{M_V^2} \right) \right] \!+\! \mathcal{O}\!\left(\!\frac{\widetilde{F}^4_{V,A}}{F^4_{V,A}}\!\right) 
. \label{tildedelta2_bosonic_approachB}
\end{align}
\end{widetext}

In case of considering both WSRs, these results and (\ref{eq.1+2WSR}) allow us to obtain the contributions from the bosonic cuts in Approach B~\cite{QCD23},
\begin{align} 
& \left.S_{\mathrm{NLO}}\right|_{\varphi\varphi, h\varphi}^{\mathrm{\bf{B}}}  \!=\!
\left.S_{\mathrm{NLO}}\right|_{\varphi\varphi, h\varphi}^{\mathrm{\bf{A}}} \!+\! 
\frac{1}{12\pi}  \left( \frac{M_A^2}{M_V^2} \!-\! 1\right)  \log \frac{M_A^2}{M_V^2} \times\nonumber \\
&\qquad  \left(\frac{\widetilde{F}_V^2}{F_V^2} +2\kappa_W^2 \frac{\widetilde{F}_V \widetilde{F}_A}{F_V F_A} - \kappa_W^2 \frac{\widetilde{F}_A^2}{F_A^2} \right)  + \mathcal{O}\!\left(\!\frac{\widetilde{F}^4_{V,A}}{F^4_{V,A}}\!\right) 
,  \label{SNLOBB}
\end{align}
where we follow the notation of (\ref{SNLOA_1}) and (\ref{SNLOA}), so that in (\ref{SNLOBB}) the contributions coming from (\ref{delta1_bosonic_approachB}) and (\ref{delta2_bosonic_approachB})  are included and the only additional ingredient would be the vanishing of (\ref{tildedelta2_bosonic_approachB}) imposed by the $2^{\mathrm{nd}}$ WSR. As it can be observed, there are many cancellations between (\ref{S_1WSR_B}) and (\ref{delta1_bosonic_approachB})-(\ref{delta2_bosonic_approachB}).

In case of considering only the $1^{\text{st}}$ WSR, the results of (\ref{S_1WSR_B}) and (\ref{delta1_bosonic_approachB}) together with (\ref{eq.NLO-S+1WSR}) allow us to obtain the lower bound (\ref{S1WSR_A_1}) for $S$ (considering only bosonic cuts) 
with~\cite{QCD23}
\begin{align} 
  &  \left.\Delta S_{\mathrm{NLO}}\right|_{\varphi\varphi, h\varphi}^{\mathrm{\bf{B}}} = \left.\Delta S_{\mathrm{NLO}}\right|_{\varphi\varphi, h\varphi}^{\mathrm{\bf{A}}}+
    \frac{1}{12\pi} \left\{ \left(1-\frac{M_A^2}{M_V^2}\right) \times \right. \nonumber \\
    &\quad \quad  \left.\left. 
    \left[ \frac{\widetilde{F}_V^2}{F_V^2}+ \kappa_W^2 \frac{\widetilde{F}_A}{F_A}\left( 2 \frac{\widetilde{F}_V}{F_V}-\frac{\widetilde{F}_A}{F_A} \right)\right]\right. \right.  \nonumber \\
    & \quad \left. \!+\! \log \frac{M_A^2}{M_V^2}\left( \frac{\widetilde{F}_V^2}{F_V^2}\!-\! \kappa_W^2 \frac{\widetilde{F}_A^2}{F_A^2} \!-\! 2 \frac{\widetilde{F}_V\widetilde{F}_A}{F_VF_A}\right)
    \right\} 
    \!+\! \mathcal{O}\!\left(\!\frac{\widetilde{F}^4_{V,A}}{F^4_{V,A}}\!\right) \label{DeltaS_B}
    ,
\end{align}
where we follow the notation of (\ref{S1WSR_A_1}) and (\ref{S1WSR_A}), so that in this result the contribution coming from (\ref{delta1_bosonic_approachB}) is included.

\subsubsection{Contributions to $T$}

The self-energy of the charged Goldstone receives a non-zero contribution from loops with a $B$ gauge boson and a Goldstone, while the contributions to the neutral self-energy originate in a $hB$ cut. The calculation of these diagrams involves the same vertices that have been used before for the $S$ parameter. Therefore, the one-loop self-energies can be also expressed in terms of the previous form factors: 
\begin{align}
\left.\Sigma^{(+)}(q^2)\right|_{\varphi B}\! \!=&\, g'^{\, 2} q_\mu q_\nu  \! \Int\!  \Frac{{\rm d^Dk}}{i(2\pi)^D}\,
   \! \left|\mF^V_{\varphi\varphi}(k^2)+ \mF^A_{\varphi\varphi}(k^2)\right|^2 \!  
   \nn \\ & \qquad \qquad \times 
\Frac{ g^{\mu\nu}-k^\mu k^\nu /k^2 }{k^2\, (q-k)^2}\, ,
\nn\\
\left.\Sigma^{(0)}(q^2)\right|_{h B}\! \!=&\,  g'^{\, 2} q_\mu q_\nu
\! \Int\!  \Frac{{\rm d^Dk}}{i(2\pi)^D}\,
   \!\left|\mF^V_{h\varphi}(k^2)+\mF^A_{h\varphi}(k^2)\right|^2\!   
 \nn \\ & \qquad\qquad \times
\Frac{ g^{\mu\nu}-k^\mu k^\nu /k^2 }{k^2  \,
  [(q-k)^2-m_{h}^2]     }
\, ,
\label{eq.Sigma-Bpi}
\end{align}
which allow us to get $T$ by using Eqs.~(\ref{calculation_T_1})-(\ref{calculation_T_3}).
Notice that the relevant form-factor combination for the $\varphi B$ and $h B$ absorptive cuts is of the form $\mF^V+\mF^A$, as the intermediate $B$ boson interacts through a $V+A$ current. 

Thus, the same $\varphi\varphi$ and $h\varphi$ form factors entering the calculation of $S$ determine the one-loop contributions to $T$. Therefore, once the conditions (\ref{constraints_FF}) have been implemented, the four form factors are very well behaved at high energies, implying also a good UV convergence of the Goldstone self-energies. This allows us to perform an unambiguous determination of $\rho_T(s)$ in terms of the resonance masses and $\kappa_W$:
\begin{widetext}   
\begin{align}
&\rho_T(s)\big|^{\mathrm{\bf{A}}}_{\varphi\varphi, h\varphi} \!= \!\Frac{g'^{\, 2}s}{64\pi^2}  \!\!\left\{ \theta(s) \!\left[\! 3(\kappa_W^2\!-\!1) \!+\! \frac{2s}{M_V^2} \!-\! \frac{2\kappa_W^2s}{M_A^2}\! \right]  \!-\! \theta(s\!-\!M_V^2)  \!\left(\!1 \!-\! \frac{M_V^2}{s}\!\right)^2\! \!\left(\! 1 \!+\! \frac{2s}{M_V^2}\! \right) 
\!+\! \theta(s\!-\!M_A^2) \kappa_W^2 \!\left(\!1 \!-\! \frac{M_A^2}{s}\!\right)^2\!\! \left(\! 1 \!+\! \frac{2s}{M_A^2}\! \right)\! \!\right\}\!, \nn \\ 
&  \rho_T(s)\big|^{\mathrm{\bf{B}}}_{\varphi\varphi, h\varphi}\!=\! \rho_T(s)\big|^{\mathrm{\bf{A}}}_{\varphi\varphi, h\varphi} + \!\Frac{g'^{\, 2}}{64\pi^2 M_A^2 M_V^2\left(M_A^2-M_V^2\right) s} \left\{ 
-2s^3 \left(M_A^2-M_V^2\right)^2 \theta(s)   \left[\frac{\widetilde{F}_V}{F_V} + \kappa_W^2 \frac{\widetilde{F}_A}{F_A} \right] 
\right.
\nn  
\\
&   \left. 
+ 2M_A^2 \left(M_V^2 \!-\!s \right)^2 \theta (s\!-\!M_V^2)\!
\left[ \frac{\widetilde{F}_V}{F_V} \Big( M_A^2 \left( s + 2M_V^2\right) -2s M_V^2 -M_V^4 \Big) + \frac{\widetilde{F}_A}{F_A} \kappa_W^2 M_A^2 \left( s-M_V^2 \right) \right] \right.
\nn 
\\
&   
+ 2M_V^2 \left(M_A^2 \!-\!s \right)^2 \theta (s\!-\!M_A^2)\!
\left[ \frac{\widetilde{F}_V}{F_V} M_V^2 \left(s-M_A^2 \right) + \frac{\widetilde{F}_A}{F_A} \kappa_W^2 \Big( s M_V^2 -M_A^4 +2M_A^2 \left( M_V^2-s\right) 
\Big) \right] 
\nn  
\\
&+  
2s^3 \left(M_A^2-M_V^2\right)^2 \left( 1 + \kappa_W^2 \right) \theta(s)    \frac{\widetilde{F}_V}{F_V} \frac{\widetilde{F}_A}{F_A}
   + M_A^2 \left(M_V^2-s\right)^2 \theta (s\!-\!M_V^2) 
\left[ \!
\kappa_W^2 \frac{\widetilde{F}_A}{F_A} \!\Bigg(\! M_A^2 \!\left(\! M_V^2 \!\left(\!2 \frac{\widetilde{F}_V}{F_V} \!+\!  3 \frac{\widetilde{F}_A}{F_A} \!\right)\!-\!2s   \frac{\widetilde{F}_V}{F_V} \!\right)  \right. 
\nn   \\
&   \left.  \!-\! M_V^2 \!\left(\!M_V^2 \!+\!2s\right)\! \frac{\widetilde{F}_A}{F_A}\! \Bigg)\!
\!  
+\! \frac{\widetilde{F}_V}{F_V}\! \Bigg(\!
M_V^2 \!\left(\!M_V^2 \!+\!2s\! \right)\! \left(\! \frac{\widetilde{F}_V}{F_V} \!+\! 2 \frac{\widetilde{F}_A}{F_A}\!\right) \!-\!M_A^2 \!\left( \!2s \frac{\widetilde{F}_A}{F_A} \!+\! M_V^2\! \left(\! 3 \frac{\widetilde{F}_V}{F_V} \!+\! 4 \frac{\widetilde{F}_A}{F_A}\! \right)\!\right)\!
\Bigg)\!
\right]\!
\nn 
\\
&  + M_V^2 \left(M_A^2-s\right)^2 \theta (s\!-\!M_A^2) \left[ 
-\frac{\widetilde{F}_V}{F_V} \Bigg( 2M_V^2 s \frac{\widetilde{F}_A}{F_A} + M_A^4 \frac{\widetilde{F}_V}{F_V} + M_A^2 \left( 2 s \frac{\widetilde{F}_V}{F_V} - M_V^2 \left( 3  \frac{\widetilde{F}_V}{F_V} + 2 \frac{\widetilde{F}_A}{F_A}  \right)\right)  \Bigg)
\right.
\nn   \\
&  \left. \left. + \kappa_W^2  \frac{\widetilde{F}_A}{F_A} \Bigg( -2 M_V^2 s \frac{\widetilde{F}_V}{F_V} + M_A^4 \left(2\frac{\widetilde{F}_V}{F_V} + \frac{\widetilde{F}_A}{F_A} \right) + M_A^2 \left( 2s \left(2\frac{\widetilde{F}_V}{F_V} + \frac{\widetilde{F}_A}{F_A} \right)-M_V^2 \left(4\frac{\widetilde{F}_V}{F_V} + 3 \frac{\widetilde{F}_A}{F_A} \right)\right)\Bigg) \right] \right\} \!+\! \mathcal{O}\!\left(\!\frac{\widetilde{F}^3_{V,A}}{F^3_{V,A}}\!\right) ,\!  \label{eq:rhoT} 
\end{align}
where terms of $\cO(m_h^2/s)$ have been neglected and we give the result in both approaches, A and B. 
Note that $\rho_T(s)\big|_{\varphi\varphi, h\varphi}=0$ in the SM up to $\mO(m_h^2/s)$ corrections. 
Following the procedure of Subsection~\ref{dispersive}, $T$ reads
\begin{align}
&T\big|^{\mathrm{\bf{A}}}_{\varphi\varphi, h\varphi}  \!=\! \frac{3}{16\pi \cos^2 \theta_W} \!\left[ \!\left( 1\!-\!\kappa_W^2 \right) \!\left( \!1 \!-\! \log{\frac{M_V^2}{m_{h}^2}} \!\right)  \!+\! \kappa_W^2 \log{\frac{M_A^2}{M_V^2}}   \right] \!  , \label{TapproachA}  
\end{align} 
and~\cite{QCD23}
\begin{align}
& T\big|^{\mathrm{\bf{B}}}_{\varphi\varphi, h\varphi}\! = \! T\big|^{\mathrm{\bf{A}}}_{\varphi\varphi, h\varphi} \!+\! \frac{3}{16\pi \cos^2 \theta_W}\! 
\Bigg\{ \! 2\kappa_W^2 \! \frac{\widetilde{F}_A}{F_A}\!-\! 2\frac{\widetilde{F}_V}{F_V} \!+\!  \frac{M_V^2}{M_A^2\!-\!M_V^2} 
 \log \frac{M_A^2}{M_V^2} \left( 2\frac{\widetilde{F}_V}{F_V} \!-\! 2 \kappa_W^2 \frac{M_A^2}{M_V^2} \frac{\widetilde{F}_A}{F_A}\right)    \nn \\ & 
 \!+\!  \frac{M_V^2}{M_A^2\!-\!M_V^2} 
 \log \frac{M_A^2}{M_V^2} \left[\,  
 \left(\kappa_W^2 \frac{\widetilde{F}_A^2}{F_A^2} \!-\! \frac{\widetilde{F}_V^2}{F_V^2}  \right)
 \left(1\!+\! \frac{M_A^2}{M_V^2} \right) 
 +2 \,\frac{ \widetilde{F}_V\widetilde{F}_A}{ F_VF_A } \left( \kappa_W^2 \frac{M_A^2}{M_V^2} \!-\! 1 \right) 
  \right]\nonumber \\ &
    \qquad\qquad +2 \!\left(
    \frac{\widetilde{F}_V^2}{F_V^2} \!-\!
    \!\kappa_W^2 \frac{\widetilde{F}_A^2}{F_A^2}  \!+\! \left( 1 \!-\! \kappa_W^2 \right)\!\frac{ \widetilde{F}_V\widetilde{F}_A}{ F_VF_A } \!\right)\! \Bigg\}
 \!+\! \mathcal{O}\!\left(\!\frac{\widetilde{F}^3_{V,A}}{F^3_{V,A}}\!\right)\!
.\label{TapproachB}  
\end{align}
\end{widetext}
As before, terms of $\mathcal{O}(m_h^2/M_{V,A}^2)$ have been neglected.
As expected, Approach A recovers the result in Ref.~\cite{ST}. 
The terms in the first line of Eq.~(\ref{TapproachB}), after $T\big|^{\mathrm{\bf{A}}}_{\varphi\varphi, h\varphi}$, provide the first-order correction ($\mO(\widetilde{F}_R/F_R)$) to the P-even limit while the second and third lines provide the second-order correction  ($\mO(\widetilde{F}_R^2/F_R^2)$).  
In order to extract the results in (\ref{TapproachA}) and~(\ref{TapproachB}) from the spectral functions in~(\ref{eq:rhoT}) a clarification is needed: we have provided the $\rho_T(s)$ functions in the limit $m_h^2/M_R^2\to 0$, neglecting corrections proportional to $m_h^2$. 
However, as the $h B$ threshold goes then down to $s=0$, the $T$--dispersive integral from (\ref{eq:T-disp-rel}) has now an infrared logarithmic divergence. Its regularization and connection with the physical $T$ (up to $\mO(m_h^2/M_{V,A}^2)$ corrections) is discussed in Appendix~\ref{App:IR-regulariza}. From a practical point of view, this procedure amounts to integrate 
from $s=m_h^2$ up to $s\rightarrow \infty$ the $hB$ contribution to the
dispersive expression in (\ref{eq:T-disp-rel}), neglecting $\mO(m_h^2/M_{V,A}^2)$ contributions.

\subsection{$\psi \bar{\psi}$ cut}

For the sake of clarity, some of the technicalities of the calculation of the fermion-antifermion ($\psi\bar\psi$) contribution to the spectral function have been relegated to Appendix~B. Neglecting the masses of SM-particles ($m_\psi=0$) and discarding also subleading contributions in $g'$, the result can be given in terms of two form factors:
\begin{eqnarray}
\rho_S(s)|_{\psi\bar\psi}& =&  \sum_{\psi} \theta(s)\; \Frac{g^2\tan\theta_W}{192\pi^2 }
\bigg[8 \,T^3_\psi\, x_\psi + \nonumber \\ &&  \left. \quad \qquad \quad 
\frac{s}{2} \!\left( \left|\mathcal{F}^{\mathcal{V}_3}_{2, \psi\bar{\psi}} \right|^2 \!-\!  \left|\mathcal{F}^{\mathcal{A}_3}_{2, \psi\bar{\psi}} \right|^2 
\right) \!
\right] , \label{rho_psipsi}
\end{eqnarray}
where all the non-SM pieces are contained in the corresponding vector and axial-vector fermion-antifermion form factors, which are reported in Appendix~B,
\begin{eqnarray}
\mathcal{F}^{\mathcal{V}_3}_{2, \psi\bar{\psi}}
&=&     \, -\,   4\sqrt{2} \, T^3_\psi\, 
\left( \frac{F_V C_0^{V_3^1}  }{M_{V}^2 - s} 
+\frac{\widetilde{F}_A \widetilde{C}_0^{A_3^1}  }{M_{A}^2 - s} \right)    \, , 
\nonumber \\
\mathcal{F}^{\mathcal{A}_3}_{2, \psi\bar{\psi}}
&=&       4\sqrt{2} \, T^3_\psi\, 
\left( \frac{\widetilde{F}_V C_0^{V_3^1}  }{M_{V}^2 - s} 
+\frac{F_A \widetilde{C}_0^{A_3^1}  }{M_{A}^2 - s} \right) \,.
\end{eqnarray}
Note that the SM contribution to Eq.~(\ref{rho_psipsi}) is consistent with (\ref{rho_SM}), once $m_\psi \rightarrow 0$ and the resonance couplings are set to zero.

The fermionic-cut contributions in (\ref{rho_psipsi}) exhibit a different high-energy behavior than the bosonic contributions in (\ref{rho_phiphi}) and (\ref{rho_hphi}). Leaving aside the SM term in (\ref{rho_psipsi}), which cancels at short distances when adding the contributions of the fermions of every family, the $\psi\bar{\psi}$ cut generates an $\cO(1/s)$ contribution to $\rho_S(s)$ while the $\varphi\varphi$ and $h\varphi$ spectral amplitudes were nominally of $\mO(s^0)$.\footnote{
This is not a surprise and is consistent with the power counting adopted in Refs.~\cite{lagrangian,lagrangian_color}. Owing to their weak coupling to the strong sector, the fermion bilinears 
are assumed to be $\cO(p^2)$, while a na\"ive dimensional analysis would assign them an $\cO(p)$ scaling.
}
In order to recover a proper UV behaviour, the $\varphi\varphi$ and $h\varphi$ form factors have been enforced before to vanish at large momenta, which leads to $\rho_S (s)|_{\varphi\varphi,h\varphi}\sim\cO(1/s^2)$. Once this is implemented, 
the contribution from the two-fermion cut would dominate at high energies, generating an $\cO(1/s)$  behaviour that is not compatible with the first WSR.

Indeed, and contrary to what happens with the $\varphi\varphi$ and $h\varphi$ cuts, the fermionic cuts generate a logarithmic contribution of $\mO(s^{-1})$ to the renormalized one-loop correlator,  
$\overline{\Pi}(s)\sim   \widetilde{\delta}^{(1)}_{_{\rm NLO}}\, \frac{v^2}{s}\log{\left(\frac{-s}{M_V^2}\right)}$, with
\begin{eqnarray}
&&\widetilde{\delta}^{(1)}_{_{\rm NLO}}|_{\psi\overline{\psi}} 
= \, -\, \frac{1}{12\pi^2v^2} \times 
\nonumber\\
&&\qquad\qquad  
\left[\left( F_V C_0^{V_3^1}    
\!+\! \widetilde{F}_A \widetilde{C}_0^{A_3^1}   \right)^2 \!-\! \left( \widetilde{F}_V C_0^{V_3^1}  
\!+\! F_A \widetilde{C}_0^{A_3^1}  \right)^2\right].
\nonumber\\
\label{eq:delta1tilde_fermions}
\end{eqnarray}

Thus, fermion cuts yield the only contribution to $\widetilde{\delta}^{(1)}_{_{\rm NLO}}$. The requirement that the 1st WSR must be fulfilled demands that the combination of resonance couplings in~(\ref{eq:delta1tilde_fermions}) vanishes. Before making any P-odd expansion, this implies
 \begin{equation}
\widetilde{C}_0^{A_3^1} \, =\,  
\mp \left( \frac{F_V\pm \widetilde{F}_V}{F_A \pm  \widetilde{F}_A}\right)\, C_0^{V_3^1}\, . 
\label{eq:delta2t-CV-CAtilde}
\end{equation}

We will further assume a theory close to the P--symmetric case, where 
$\widetilde{F}_V/F_V$ and $\widetilde{F}_A/F_A$    
are of a similar symmetry-breaking order that we denote as $\mO(\epsilon_P)$. 
A thorough analysis shows that, in addition to the trivial solution $C_0^{V^1_3}= \widetilde{C}_0^{A_3^1}=0$, 
the fulfilment of the identity~(\ref{eq:delta2t-CV-CAtilde}) requires that both $C_0^{V_3^1}$ and $\widetilde{C}_0^{A_3^1}$
have a similar suppression of $\mO(\epsilon_P)$. Expanding around the P--symmetric limit, one finds:
\begin{eqnarray}
\left( F_V C_0^{V_3^1}\right)^2 \,=\, \left( F_A \widetilde{C}_0^{A_3^1}\right)^2 \, + \, \mO(\epsilon_P^3) \, ,
\end{eqnarray} 
which implies, 
\begin{eqnarray}
\left|\mathcal{F}^{\mathcal{V}_3}_{2, \psi\bar{\psi}} \right|^2 \!-\!  \left|\mathcal{F}^{\mathcal{A}_3}_{2, \psi\bar{\psi}} \right|^2 & =& 8\, \left( F_V C_0^{V_3^1}\right)^2
\left[ \frac{1}{(M_V^2-s)^2}\right.
\nonumber\\
&& \left. -\, \frac{1}{(M_A^2-s)^2}\right] \, +\, \mO(\epsilon_P^3)\,,
\end{eqnarray}
 where the lowest non-trivial contribution is of $\mO(\epsilon_P^2)$.

Neglecting terms of $\mO(\epsilon_P^3)$, one finally finds the following contributions to the $S$--parameter:
\begin{eqnarray}
\delta_{_{\rm NLO}}^{(1)}\big|_{\psi\bar\psi} &=& - \, \frac{F_V^2\, ( C_0^{V^1_3})^2 }{12\pi^2 v^2} \log{\frac{M_A^2}{M_V^2}}\,,
\nonumber\\
\delta_{_{\rm NLO}}^{(2)}\big|_{\psi\bar\psi} &=&\frac{F_V^2\, ( C_0^{V^1_3})^2 }{12\pi^2 v^2} \left(1\, -\, \frac{M_A^2}{M_V^2} -\frac{2 M_A^2}{M_V^2}\log{\frac{M_A^2}{M_V^2}}\right)\,, 
\nonumber\\
\widetilde\delta_{_{\rm NLO}}^{(2)}\big|_{\psi\bar\psi} &=& \frac{F_V^2\, ( C_0^{V^1_3})^2}{6\pi^2 v^2} \left(\frac{M_A^2}{M_V^2}-1\right)\,,
\nonumber\\
\left.\overline{S}\right|_{\psi\bar\psi} &=& 
  - \frac{F_V^2\, ( C_0^{V^1_3})^2 }{3\pi M_V^2}\left(1-\frac{M_V^2}{M_A^2}\right) \,.   
\end{eqnarray}
Combining these results, the total fermionic contributions to the $S$--parameter take the forms (up to $\mO(\epsilon_P^3)$ corrections):
\begin{eqnarray}\label{eq:SNLOfermion}
\left.S_{\mathrm{NLO}}\right|_{\psi\bar\psi} &=&   \frac{F_V^2\, ( C_0^{V^1_3})^2 }{3\pi M_V^2}\left(1-\frac{M_V^2}{M_A^2}\right) \log{\frac{M_A^2}{M_V^2}} \,,   
\\ \label{eq:SNLOfermion1WSR}
\left.\Delta S_{\mathrm{NLO}}\right|_{\psi\bar\psi} &=&   -\, \frac{F_V^2\, ( C_0^{V^1_3})^2 }{3\pi M_V^2}\left(1-\frac{M_V^2}{M_A^2}    
+ \log{\frac{M_A^2}{M_V^2}}  \right) \,,  \nonumber\\
\end{eqnarray}
to be inserted in ~(\ref{SNLOA_1}) and~(\ref{S1WSR_A_1}), respectively, for the two WSRs and 1st WSR cases.

Furthermore, the two WSRs of (\ref{eq:FVFA-WSRs}) provide at $\mO(\epsilon_P^0)$ the relation $F_V^2= v^2 M_A^2/(M_A^2-M_V^2)$, 
which allows us to rewrite the fermionic contribution~(\ref{eq:SNLOfermion}) in the simpler form
\begin{eqnarray}
\left.S_{\mathrm{NLO}}\right|_{\psi\bar\psi} &=&   \frac{v^2\, ( C_0^{V^1_3})^2 }{3\pi M_V^2} \log{\frac{M_A^2}{M_V^2}} 
+ \mO(\epsilon_P^3)\, .  
\label{eq:fermions-Snlo12}
\end{eqnarray} 
An estimate of ~(\ref{eq:SNLOfermion1WSR}), which only considers the 1st WSR, can be obtained by assuming that $F_V^2 \sim v^2$. 
Therefore, up to a logarithmic dependence on the ratio $M_A^2/M_V^2$, the absolute size of the fermion-cut contributions is roughly bounded by the ratio $v^2  ( C_0^{V^1_3})^2  /(3\pi M_V^2)$, both in~(\ref{eq:SNLOfermion1WSR}) and~(\ref{eq:fermions-Snlo12}).

An upper limit on the vector coupling $(C_0^{V^1_3})^2$ can be extracted
from LHC diboson-production studies ($W W$, $W Z$, $ZZ$, $W h$ and $Zh$; see~\cite{Dorigo:2018cbl} and references therein). Adapting to our more general theoretical framework the phenomenological analysis performed within the so-called Heavy-Vector-Triplet model~B~\cite{Pappadopulo:2014qza}, Ref.~\cite{lagrangian_color} obtained the (95\% CL) experimental constraint 
$ ( C_0^{V^1_3})^2 <  5 \times  10^{-3}$, for $M_V \leq  4.15$~TeV.
Assuming that for heavier vector masses this coupling does not grow faster than $M_V^2$, one finds that the fermionic contributions to the $S$ parameter happen to be extremely small. The experimentally suppressed ratio,
\begin{equation}
\frac{v^2\, ( C_0^{V^1_3})^2 }{3\pi M_V^2}\, \lsim\,   3 \cdot 10^{-5} \, ,
\end{equation}
is actually not very much enhanced by the logarithmic factors, since $\log{(M_A^2/M_V^2)}\lsim 10$ even for such a huge $V$-$A$ splitting as $M_A\sim 10^2 M_V$.  
The size of the fermion-cut contribution is essentially invisible in our plots for the oblique parameters, within the much larger uncertainties of the order of $\delta S\sim 0.1$. Hence, this contribution will be finally neglected and dropped in our analysis.

\section{Phenomenology} \label{sec:phenomenology}

\begin{figure*}
\begin{center}
\includegraphics[scale=0.6]{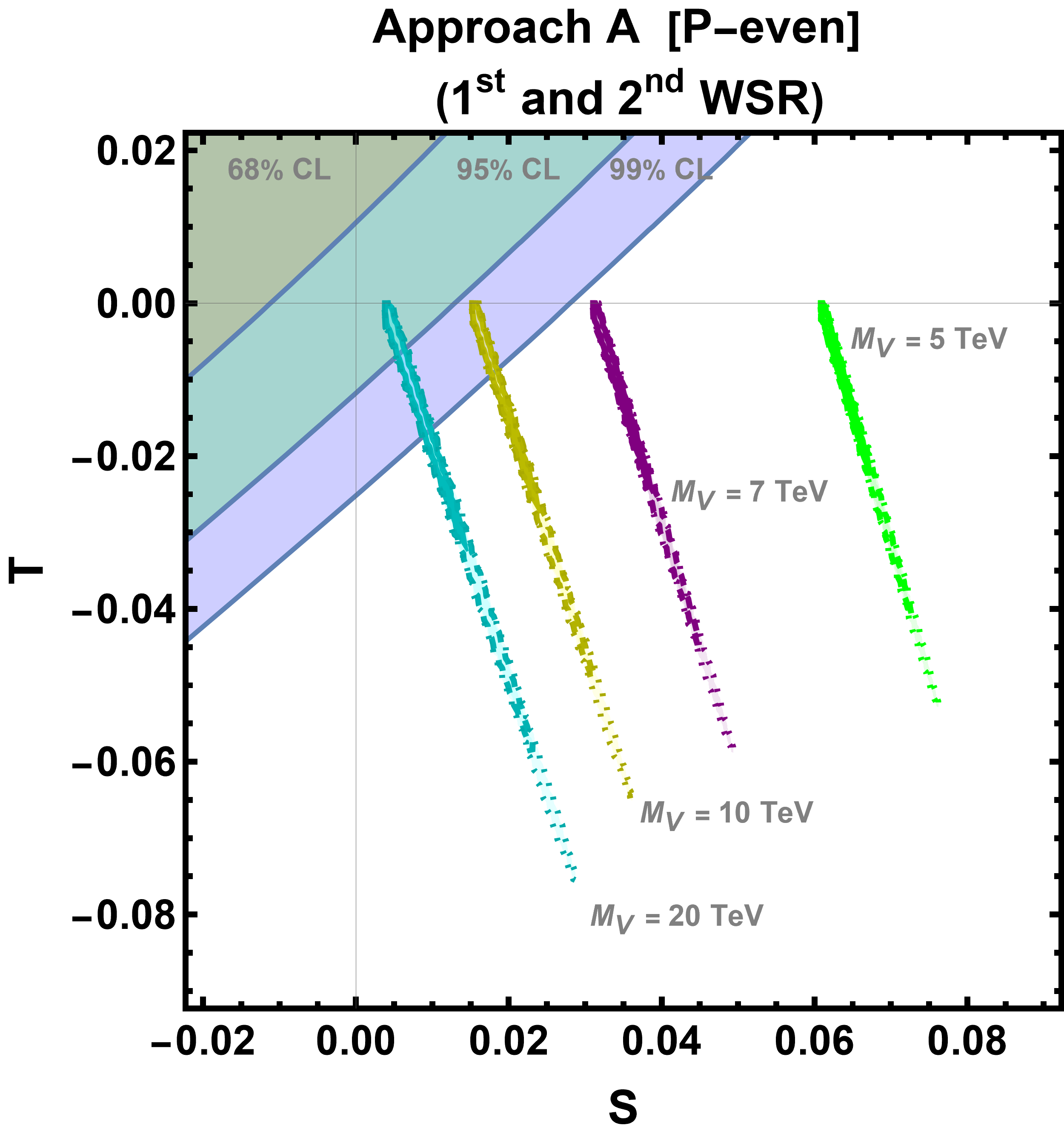} 
\hskip .75cm
\includegraphics[scale=0.6]{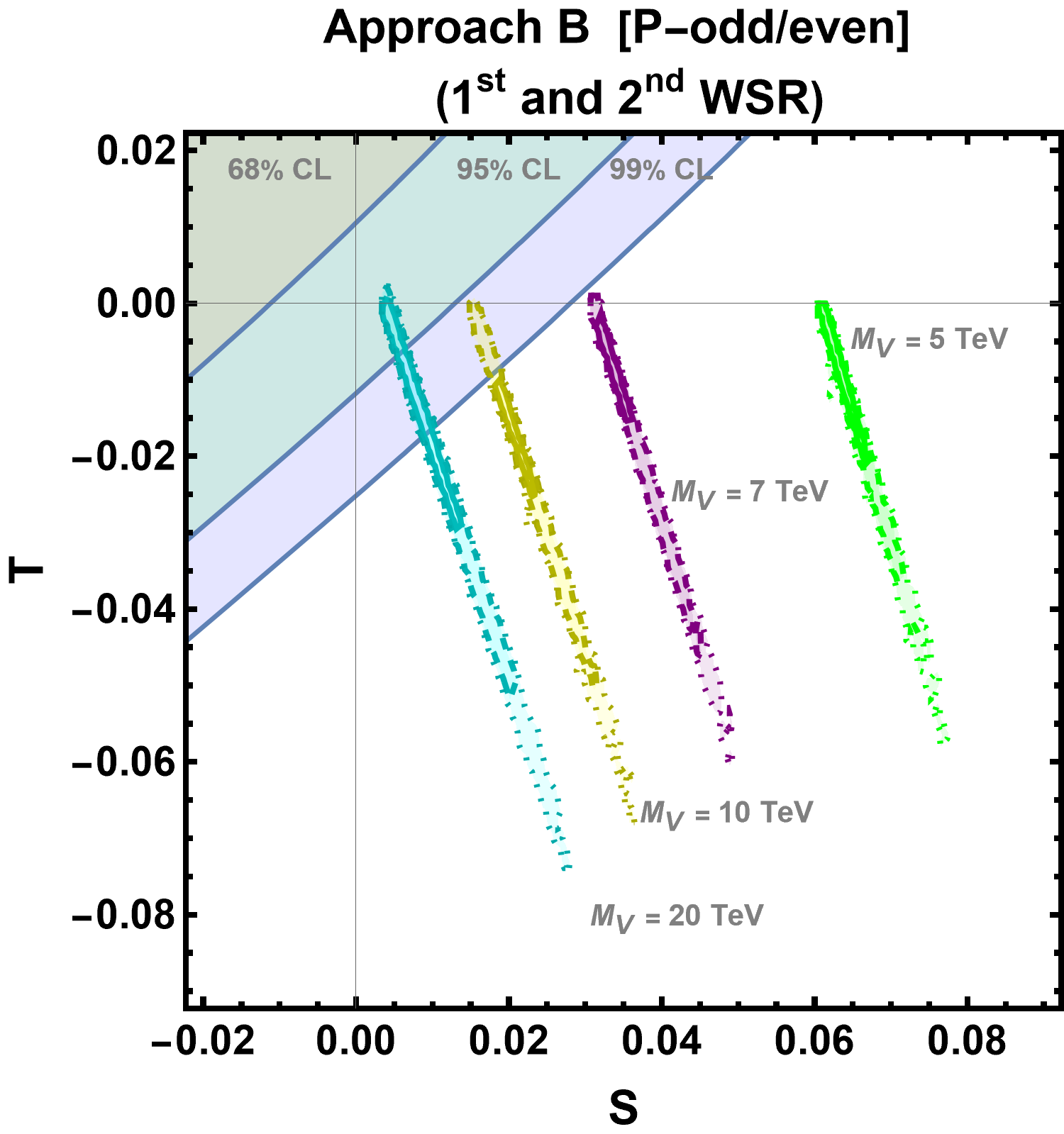} 
\caption{{\small
NLO determinations of $S$ and $T$, following Approach A (P--even; left) or B (P--odd/even; right), and assuming both WSRs. The ellipses give the experimentally allowed regions of $S$ and $T$ at $68$\%, $95$\% and $99$\% CL~\cite{PDG}. The different colors of the points correspond to different values of $M_V$: 
$M_V=5$ (green), $7$ (purple), $10$ (yellow) and $20$ (cyan) TeV. For each case, we plot our predictions at $68$\%, $95$\% and $99$\% CL.  
$M_A$ is determined by $\widetilde\delta_{_{\rm NLO}}^{(2)}=0$, taking into account that $M_A\!>\!M_V$. These constraints imply in general  values of $M_A$ very close to $M_V$, so that similar results are obtained in both approaches, A and B, respectively including only-P-even operators and both P-even and P-odd terms. 
The values of $\kappa_W$ (Approaches A and B) and $\widetilde{F}_{V,A}/F_{V,A}$ (Approach B)  have been generated considering normal distributions given by $\kappa_W=1.023\pm 0.026$~\cite{PDG} and $\widetilde{F}_{V,A}/F_{V,A}=0.00\pm 0.33$.  }}
\label{fig:NLO_2WSR}
\end{center}
\end{figure*}

\begin{figure*}
\begin{center}
\includegraphics[scale=0.46]{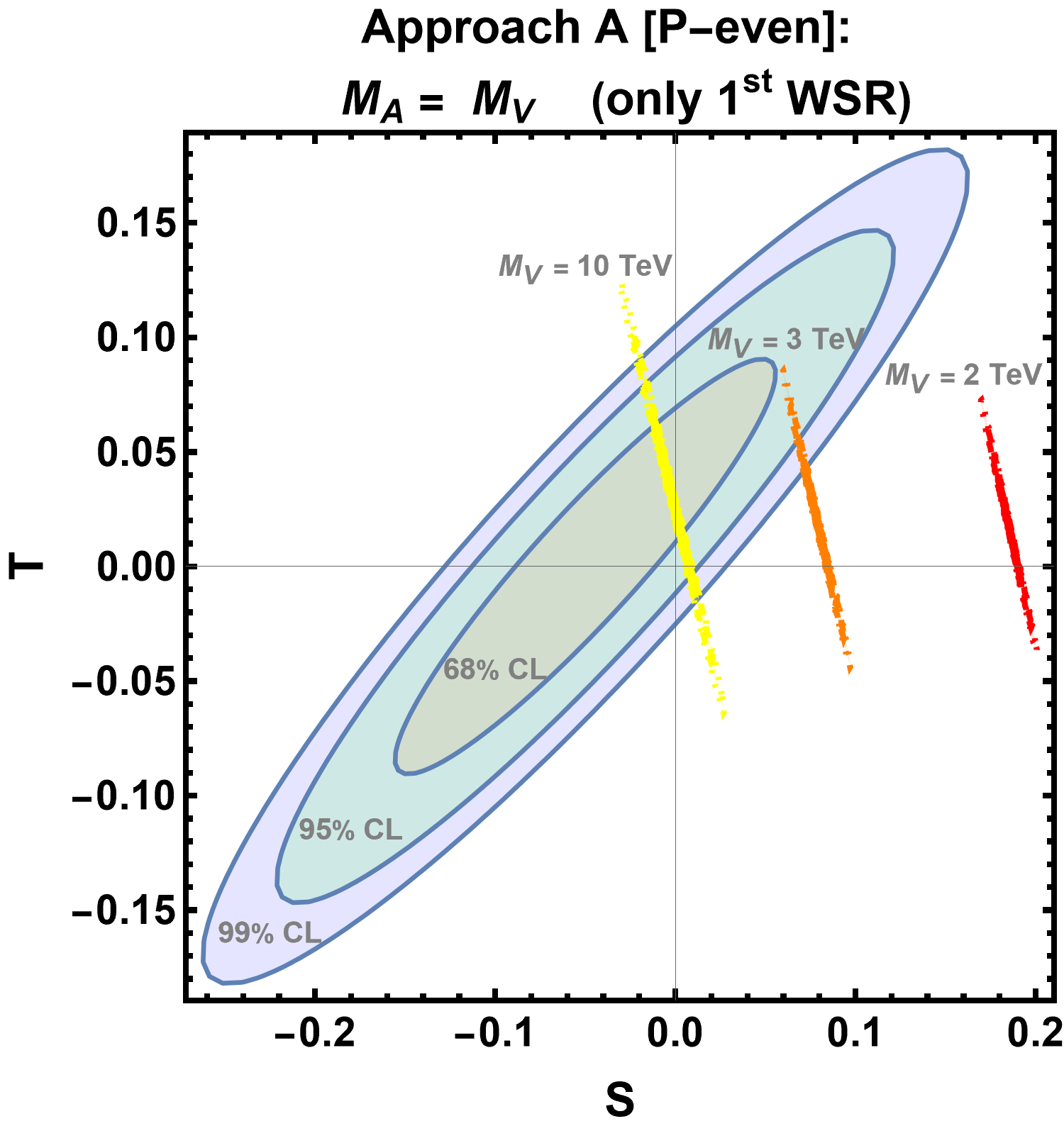} \includegraphics[scale=0.46]{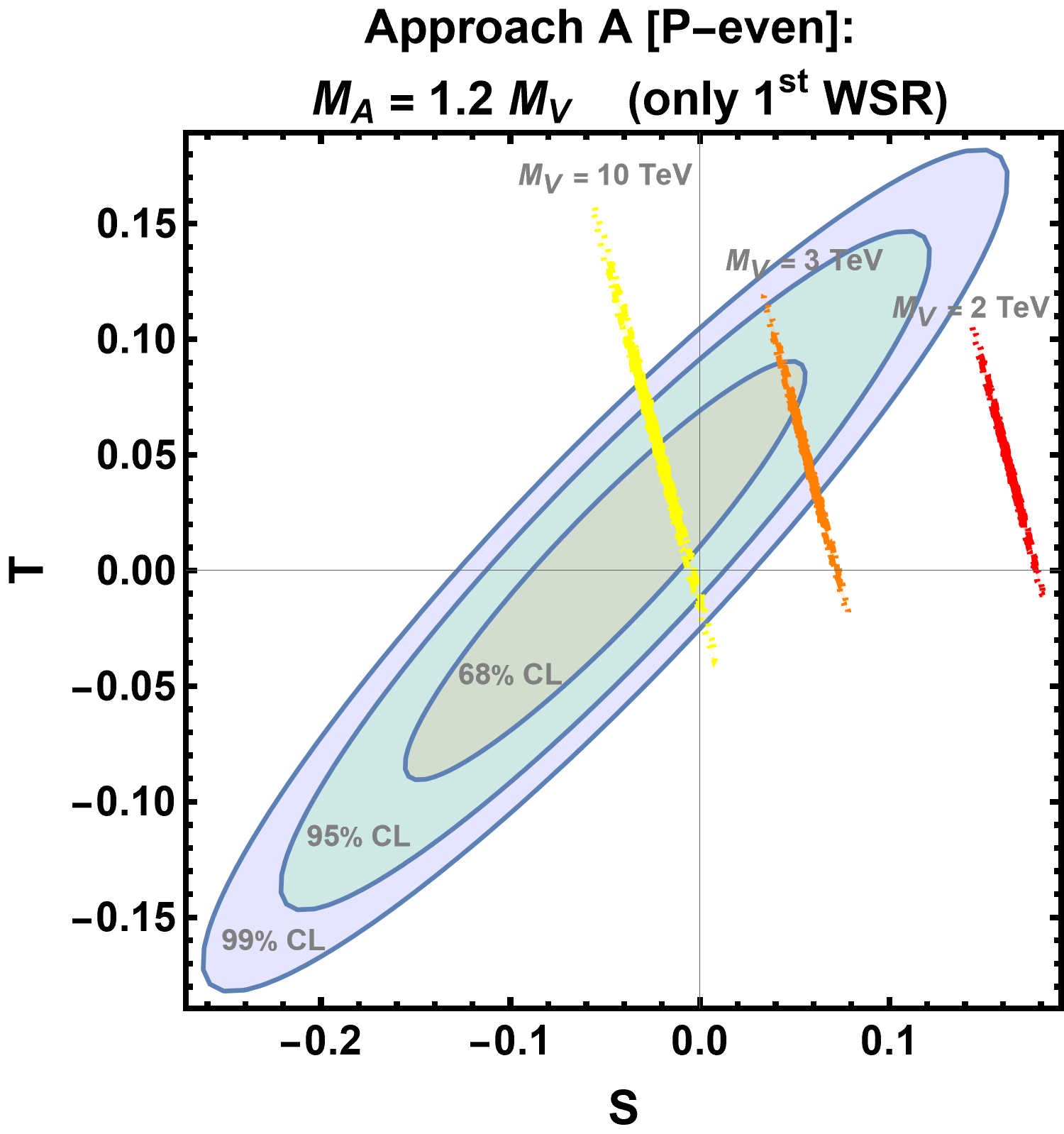} 
\includegraphics[scale=0.46]{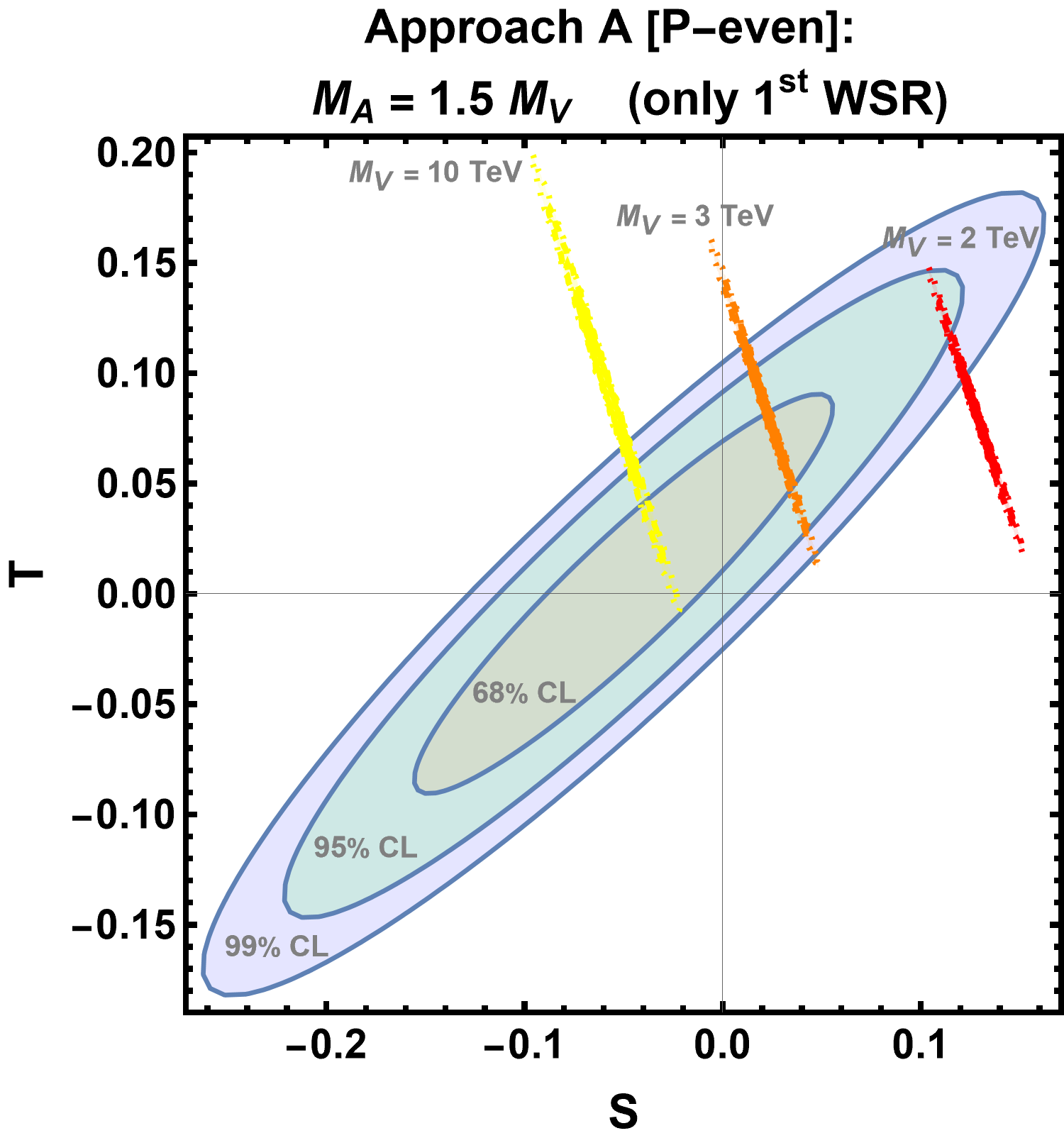} 
\\
\includegraphics[scale=0.46]{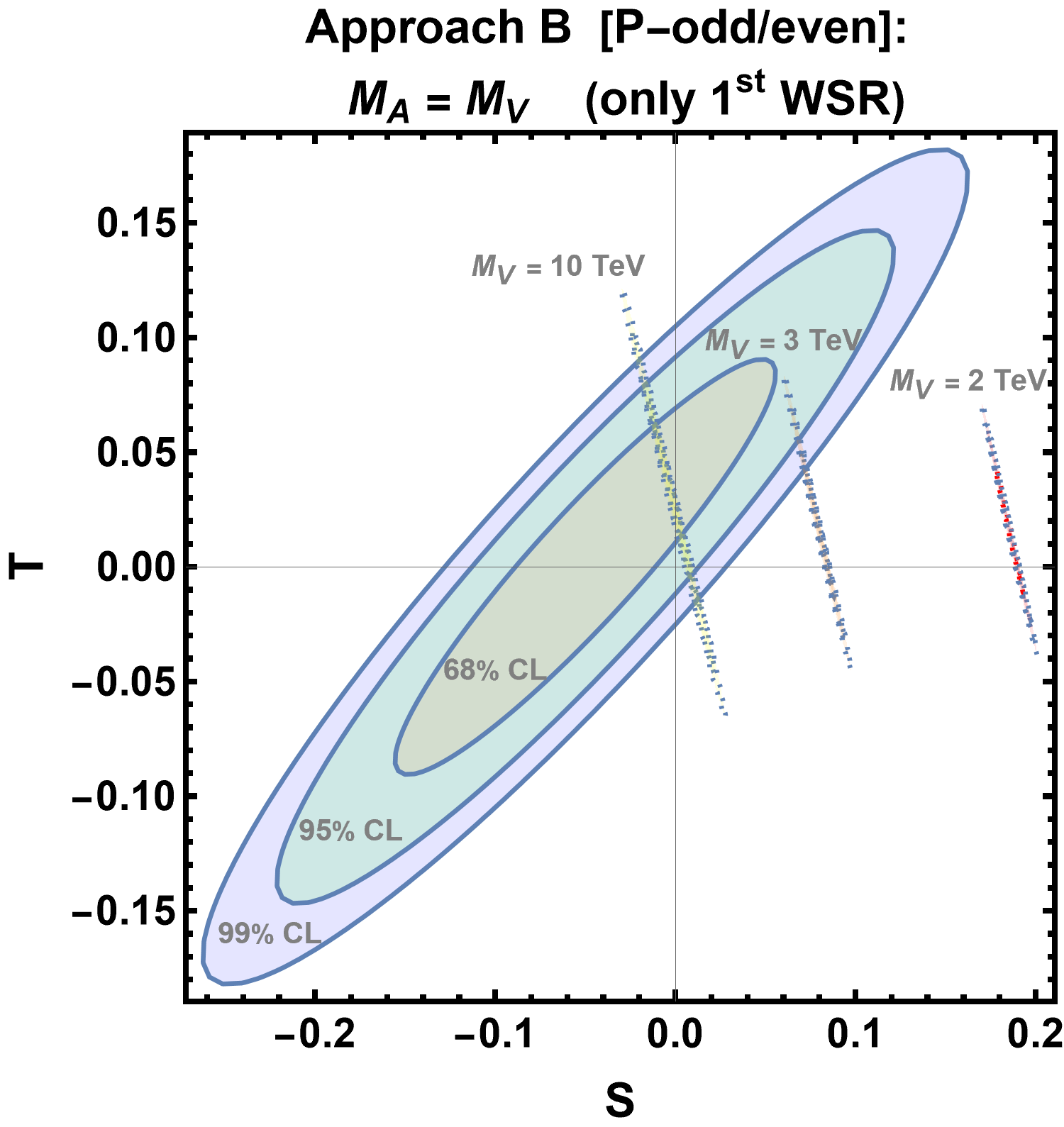} \includegraphics[scale=0.46]{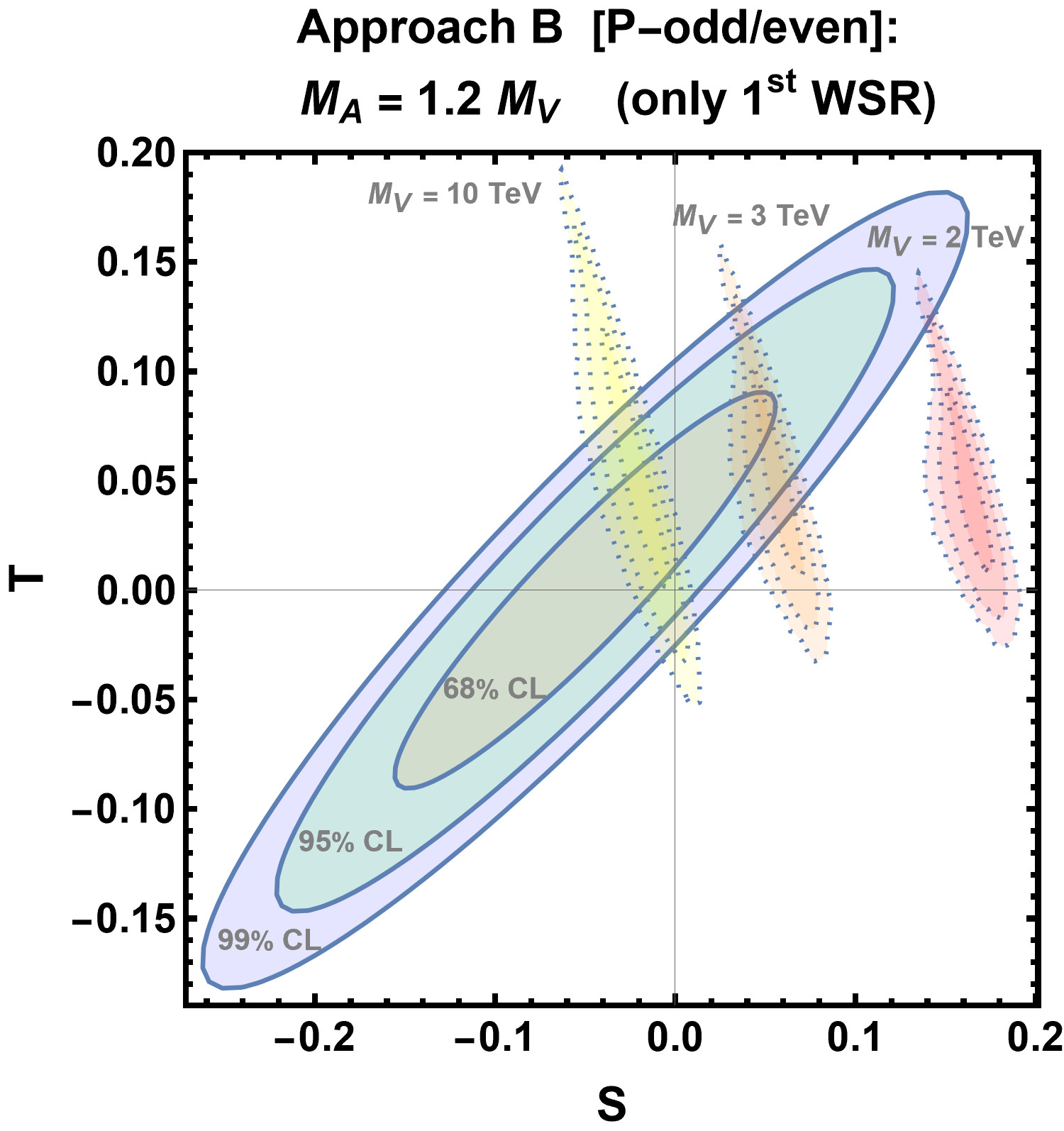} 
\includegraphics[scale=0.46]{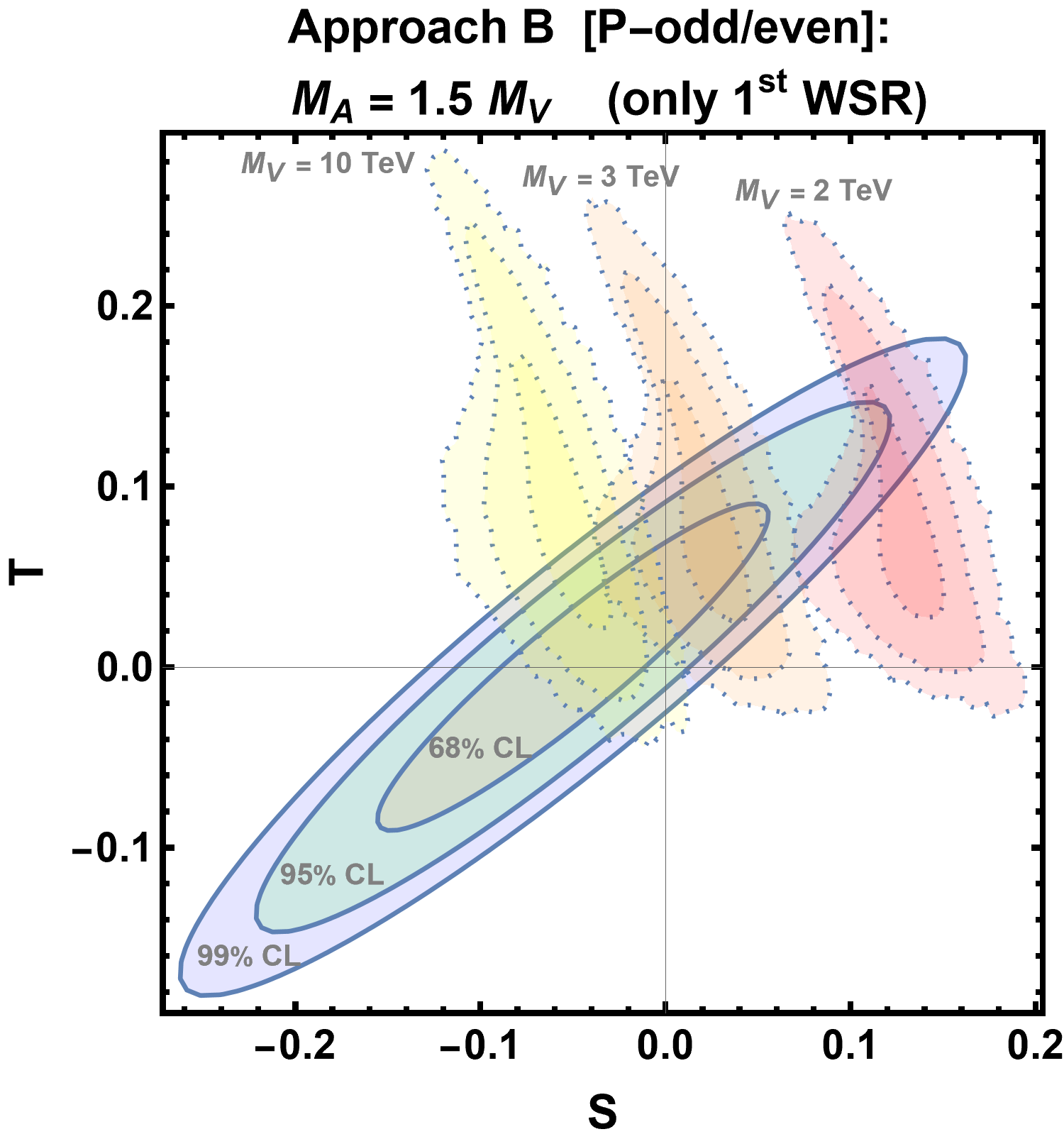}
\caption{{\small
NLO determinations of $S$ and $T$, following Approach A (P--even; top) or B (P--odd/even; bottom), and assuming only the $1^{\mathrm{st}}$ WSR, so only lower bounds on $S$ are shown. The ellipses give the experimentally allowed regions of $S$ and $T$ at $68$\%, $95$\% and $99$\% CL~\cite{PDG}. The different colors of the points correspond to different values of $M_V$: $M_V=2$ (red), $3$ (orange) and $10$ (yellow) TeV. $M_A$ is assumed to be higher than $M_V$ and we have considered different values in terms of $M_V$: $M_A=M_V$ (left), $M_A=1.2\; M_V$ (centered) and $M_A=1.5\; M_V$ (right). For each case, we plot our predictions at $68$\%, $95$\% and $99$\% CL.  
The values of $\kappa_W$ (Approaches A and B) and $\widetilde{F}_{V,A}/F_{V,A}$ (Approach B)  have been generated considering normal distributions given by $\kappa_W=1.023\pm 0.026$~\cite{PDG} and $\widetilde{F}_{V,A}/F_{V,A}=0.00\pm 0.33$. }}
\label{fig:NLO_1WSR}
\end{center}
\end{figure*}

\subsection{Approach A}

We remind again that in this Approach we are neglecting odd-parity couplings and then, taking into account that the fermionic contributions can be neglected, we recover the results of Ref.~\cite{ST}. The only difference is that, as it has been explained in Section~\ref{sec:introduction}, we no-longer consider $\kappa_W$ as a free parameter and we take instead its experimental value given in Ref.~\cite{PDG}, $\kappa_W=1.023\pm 0.026$. Within this Approach, 
$S$ and $T$
are given in terms of only two free parameters, $M_V$ and $M_A$. Depending on the assumptions related to the Weinberg Sum Rules, we consider two possibilities:
\begin{enumerate}
\item {\bf $1^{\mathrm{st}}$ WSR and $2^{\mathrm{nd}}$ WSR}. 
If both WSRs are assumed, $S$ and $T$ are determined in (\ref{SNLOA_1})-(\ref{SNLOA}) 
(neglecting the fermion-cut contribution)
and (\ref{TapproachA}), respectively. 
Furthermore, the vanishing of $\widetilde\delta_{_{\rm NLO}}^{(2)}$ can be used to determine $M_A$ in terms of $M_V$, see (\ref{tildedelta2_bosonic_approachA}):
\begin{equation} 
\widetilde\delta_{_{\rm NLO}}^{(2)}\big|^{\mathrm{\bf{A}}} 
\!=\! 0 \quad \longrightarrow \quad 
M_A^2 \!=\! \frac{M_V^2}{\kappa_W}\, , \label{MA_MV}
\end{equation}
where $M_A$ is required to be higher than $M_V$, as explained in Section~\ref{sec:WSRS}, so that $\kappa_W$ is constrained to be $\kappa_W\! <\! 1$. 
The left panel in Figure~\ref{fig:NLO_2WSR} shows the results which follow from these assumptions.
\item {\bf $1^{\mathrm{st}}$ WSR}. If one assumes only the $1^{\mathrm{st}}$ WSR and $M_A \!>\! M_V$, a lower bound of $S$ and a determination of $T$ are given in (\ref{S1WSR_A_1})-(\ref{S1WSR_A}) 
(neglecting again the fermion-cut contribution) and (\ref{TapproachA}), respectively. 
The comparison between our estimations and the experimental values is shown in the top panels of Figure~\ref{fig:NLO_1WSR}.
\end{enumerate}

\subsection{Approach B}

In this Approach odd-parity couplings are supposed to be subleading and, consequently, an expansion in $\widetilde{F}_{V,A}/F_{V,A}$ is followed. 
The expressions of $S$ and $T$ are then given in terms of four free parameters: $M_V$, $M_A$, $\widetilde{F}_{V}/F_{V}$ and $\widetilde{F}_{A}/F_{A}$, but the last two ones are expected to be small in our expansion and we will assume a normal distribution with $\widetilde{F}_{V,A}/F_{V,A}=0.00\pm 0.33$. Anew depending on the assumed Weinberg Sum Rules we have two possibilities:
\begin{enumerate}
\item {\bf $1^{\mathrm{st}}$ WSR and $2^{\mathrm{nd}}$ WSR}. If one assumes both WSRs, $S$ and $T$ are determined in (\ref{SNLOA_1}), (\ref{SNLOA}) and (\ref{SNLOBB}) (neglecting the fermion-cut contribution), and (\ref{TapproachA}) and 
(\ref{TapproachB}), respectively. 
In Section~\ref{sec:WSRS} it is demonstrated that, by considering both WSRs and within this Approach, $M_A$ is required to be higher than $M_V$.
In addition, we demand the 2nd WSR constraint  $\widetilde\delta_{_{\rm NLO}}^{(2)}  = 0  $  for the imaginary part of the loop contribution in Eqs.~(\ref{tildedelta2_bosonic_approachA}) and (\ref{tildedelta2_bosonic_approachB}).  We analytically extract $M_A$ as a function of $M_V$, $\kappa_W$, $\widetilde{F}_V/F_V$ and $\widetilde{F}_A/F_A$ and employ this value in the Approach-B predictions of Figure~\ref{fig:NLO_2WSR}.   
Expanding this solution up to $\mO(\epsilon_P^2)$ in the parity violation expansion, one can observe how Eq.~(\ref{MA_MV}) becomes now corrected:
\begin{eqnarray}
&&
M_A^2 \!=\! \frac{M_V^2}{\kappa_W}\, \left(1\, 
+\frac{(\kappa_W-1)^2}{2} \left[  \frac{\widetilde{F}_A^2}{F_A^2} -\frac{\widetilde{F}_V^2}{\kappa_W^2 F_V^2} \right] 
\right.
\label{MA_MV_epsP2}
\\
&&\qquad\qquad \left. +\frac{(\kappa_W^2-1) }{\kappa_W}   \frac{\widetilde{F}_V\widetilde{F}_A}{F_V F_A}  \, 
+\,  \mathcal{O}\!\left(\!\frac{\widetilde{F}^4_{V,A}}{F^4_{V,A}}\!\right) 
\right)\, .
\nonumber
\end{eqnarray}
We obtain very close values to $M_V$, $M_A \!\gsim\! M_V$.
In the right panel of Figure~\ref{fig:NLO_2WSR} we show the results following from these assumptions.
\item {\bf $1^{\mathrm{st}}$ WSR}. Assuming only the $1^{\mathrm{st}}$ WSR and $M_A\! >\! M_V$, we have reported a lower bound of $S$ and a determination of $T$ in (\ref{S1WSR_A_1}), (\ref{S1WSR_A}) and (\ref{DeltaS_B}) (neglecting once more the fermion-cut contribution), and (\ref{TapproachA}) and  
(\ref{TapproachB}), respectively. 
The comparison between our results and the experimental values is shown in the bottom panels of Figure~\ref{fig:NLO_1WSR}.
\end{enumerate}

\section{Discussion}\label{sec:conclusions}

Using a general (non-linear) effective field theory description of the SM EWSB, we have analysed the impact on the electroweak oblique parameters of hypothetical heavy resonance states strongly coupled to the SM particles. We have presented a next-to-leading order calculation of $S$ and $T$ that updates and generalizes our previous results in Ref.~\cite{ST}, including a more general Lagrangian~\cite{lagrangian,lagrangian_color}, fermionic cuts and the current experimental bounds~\cite{PDG}.  In particular, we have studied the numerical sensitivity to subleading contributions from P-odd operators that were neglected in Ref.~\cite{ST}.

The use of dispersion relations has avoided any dependences on unphysical cut-offs. Another important ingredient of our analysis are the high-energy constraints enforced in the effective field theory description. These are very generic conditions, which originate from requiring a proper UV behaviour of the underlying strongly-coupled theory. Assuming well-behaved form factors~\cite{PRD2} and the WSRs~\cite{WSR} allows us to determine $S$ and $T$ in terms of only a few resonance parameters. The two WSRs are rigorously fulfilled in any asymptotically-free gauge theory~\cite{Bernard:1975cd}. Gauge theories with non-trivial UV fixed points are also expected to satisfy the 1st WSR, while the validity of the 2nd WSR depends on the particular type of UV theory considered~\cite{1stWSR}. Therefore, we have performed the analyses in the two possible situations, with and without imposing the 2nd WSR, so that
our results can be applied in full generality.

At LO the oblique parameter $T$ vanishes, while $S$ only receives contributions from tree-level exchanges of vector and axial-vector resonances. The NLO corrections are dominated by the lightest two-particle cuts ($\varphi\varphi$, $h\varphi$ and $\psi\bar\psi$); contributions from multi-particle cuts involving heavy resonances have been estimated to be very small, owing to their kinematic suppression in the dispersion relation~ \cite{S_Higgsless}.

Assuming that the odd-parity couplings generate subleading corrections,
we have performed an expansion in powers of $\widetilde F_{V,A}/F_{V,A}\sim\mathcal{O}(\epsilon_P)$, and compared the lowest-order results of $\mathcal{O}(\epsilon_P^0)$ (Approach A) with those obtained at $\mathcal{O}(\epsilon_P^2)$ (Approach B). While the first Approach updates our previous work with the more recent data, the second one allows us to assess the possible role of P-odd operators.

An important finding of this analysis is that the first WSR enforces a severe suppression of the fermion-cut contribution to the $S$ parameter (contributions to $T$ are suppressed by additional powers of $g'$). The leading contribution is of $\mathcal{O}(\epsilon_P^2)$ and its size can be bounded by LHC data as being smaller than $10^{-4}$ and therefore completely negligible compared to the current experimental error of the parameter $S$.

Figure~\ref{fig:NLO_2WSR} summarizes the results of our analysis for the underlying theories that satisfy the two WSRs. This is a very constraining condition that implies $M_A \ge M_V$. Moreover, at $\mathcal{O}(\epsilon_P^0)$ (left panel) $k_W = M_V^2/M_A^2 \le 1$; the current experimental value $k_W = 1.023\pm 0.026$~\cite{PDG} forces then the vector and axial-vector masses to be quite degenerate. Those two masses remain quite close even when $\mathcal{O}(\epsilon_P^2)$ corrections are included (right panel), so that very similar results are obtained in the two Approaches, A and B. This is clearly exhibited in the figure, where one cannot see any sizeable difference between the two panels. The ellipses display the experimentally allowed regions of $S$ and $T$ at 68\%, 95\% and 99\% CL, while the colored points correspond to the predicted values for
$M_V=5$ (green), $7$ (purple), $10$ (yellow) and $20$ (cyan) TeV.   
The predictions for lighter vector masses lie outside the range displayed and are obviously excluded. Within each color, the plotted variation corresponds to the range of $M_A$ values allowed by the WSRs, and (in Approach B) the variation of the P-odd couplings in the range $\widetilde F_{V,A}/F_{V,A}= 0.00\pm 0.33$, assuming a normal distribution.  These results can be summarized in a quite strong statement: in any strongly-coupled underlying theory where the two WSRs are satisfied,
\begin{equation}
M_A\, \ge \, M_V \gsim  10\;\mathrm{TeV} \quad (95\%\; \mathrm{CL}).
\end{equation}

There is much more flexibility when the underlying theory does not satisfy the 2nd WSR because the vector and axial-vector masses are no-longer so tightly related. Assuming that the inequality $M_A > M_V$ is still fulfilled, we then obtain the results displayed in Figure~\ref{fig:NLO_1WSR}.
The colored regions show the predicted lower bounds on $S$ and the corresponding value of $T$, for $M_V=2$ (red), $3$ (orange) and $10$ (yellow) TeV.
The results are displayed for three different values of the mass ratio $M_A/M_V$: 1 (left), 1.2 (center) and 1.5 (right).
The top panels correspond to Approach A and the bottom ones to Approach B. As in the previous scenario, the distribution of points within each color has been generated considering normal distributions for $k_W = 1.023\pm 0.026$ and (in Approach B) $\widetilde F_{V,A}/F_{V,A}= 0.00\pm 0.33$. Obviously, one now gets a much broader distribution of points in Approach B, although a similar trend is observed in the two Approaches. The lower bound on $S$ decreases when $M_V$ increases, while larger values of $M_A$ imply smaller lower bounds on $S$ and slightly larger values of $T$. From these results, we can conclude that, for underlying theories where the 2nd WSR does not apply, the current electroweak precision data allow for massive resonances at the natural electroweak scale, i.e.,
\begin{equation}
M_{A}\, \geq \, M_V \, \gsim  2 \;\mathrm{TeV} \quad (95\%\; \mathrm{CL}).
\end{equation}

In summary, we conclude that the $P$-odd operators and the contributions from the fermionic cuts discussed in this article introduce mild corrections to the oblique parameters and, hence, to  our previous $M_{V,A}$ mass bounds for theories including only P-even operators.  
These findings corroborate the conclusions drawn in prior research~\cite{ST}, providing additional evidence in support of them.


\acknowledgments

A.P. would like to thank the high-energy physics group of the University of Granada for their hospitality during the time this article was being prepared. 
We thank J. Mart\'\i nez-Mart\'\i n for useful discussions. This work has been supported in part by the Spanish Government (PID2019-108655GB-I00, PID2020-114473GB-I00, PID2022-137003NB-I00, PID2023-146220NB-I00); 
financed by Spanish MCIN/AEI/10.13039/501100011033/ and FEDER programs; by EU grant 824093 (STRONG2020); by the Generalitat Valenciana (PROMETEU/2021/071); by the Universidad Cardenal Herrera-CEU (INDI24/17 and GIR24/16); by the ESI International Chair@CEU-UCH; 
by EU COMETA COST Action CA22130; by the Universidad Complutense de Madrid under research group 910309 and by the IPARCOS institute.

\appendix

\section{\boldmath $T$ and $S$ dispersion relations in the $m_h=0$ limit}
\label{App:IR-regulariza}

The parameter $T$ is given by the expression: 
\begin{eqnarray}
T&=&\frac{4\pi}{g^{' 2}\cos^2\theta_W} \int_0^\infty \frac{\Delta\rho_T(s)\, ds}{s^2}
\\
&&\hspace*{-0.65cm} =
\frac{4\pi}{g^{' 2}\cos^2\theta_W}\bigg(\! \int_0^\infty \frac{\Delta\rho_T(s)|_{\varphi B}\, ds}{s^2} 
\!+\!\int_{m_h^2}^\infty \frac{\Delta\rho_T(s)|_{h B}\, ds}{s^2}\!\bigg) ,
\nn
\end{eqnarray}
with $\Delta \rho_T(s)=\rho_T(s)-\rho_T(s)_{\rm SM}$. 
This integral is infrared (IR) divergent in the limit $m_h=0$ and needs to be regulated with an IR cut-off $\epsilon$, which defines:
\begin{eqnarray}
T^{(0)}_\epsilon &\equiv &\frac{4\pi}{g^{' 2}\cos^2\theta_W} \int_{\epsilon\to 0}^\infty \frac{\Delta\rho_T(s)\, ds}{s^2}\bigg|_{m_h=0}
\, .
\label{eq:T0eps}
\end{eqnarray}

The difference of these two expressions yields:
\begin{eqnarray}
&&T-T^{(0)}_\epsilon
\nonumber\\
&&
 \,=\,  - \frac{4\pi}{g^{' 2}\cos^2\theta_W} \!\int_{\epsilon\to 0}^{m_h^2} \frac{\Delta\rho_T(s)|_{h B,\, m_h=0}}{s^2}\, ds  \! +\! \mO\left(m_h^2\right)
\nn\\
&&\, = \, \frac{3}{16\pi\cos^2\theta_W} (\kappa_W^2-1)\log{\frac{\epsilon}{m_h^2}} \, +\, \mO\left(m_h^2\right) \, , 
\label{eq:T-Teps}
\end{eqnarray}
where we have used the structure of the spectral function at $m_h=0$,
provided by the low-energy EW effective theory at LO, $\mO(p^2)$, and neglecting $\mO(p^4)$ and higher-order corrections: 
\begin{eqnarray}
\Delta\rho_T(s)  \bigg|_{h B,\, m_h=0} &=& \frac{3g'^{\, 2} s}{64\pi^2}(\kappa_W^2-1)\, \theta(s) \, +\, \mO(s^2)\, ,   
\nn\\
\end{eqnarray}
Notice that only the $hB$--cut contributes to this expression at lowest order in the chiral expansion, i.e., $\Delta\rho_T(s) |_{m_h=0}= \Delta\rho_T(s) |_{h B,\, m_h=0}+\mO(s^2)$.

By means of Eqs.~(\ref{eq:T0eps}) and~(\ref{eq:T-Teps}) it is then possible to directly use the $m_h=0$ spectral functions in~(\ref{eq:rhoT}) to extract the parameter $T$ in Eqs.~(\ref{TapproachA}) and~(\ref{TapproachB}), up to $\mO(m_h^2/M_R^2)$ corrections.

Following a similar argumentation for the Peskin-Takeuchi dispersive relation in Eq.~(\ref{Peskin-Takeuchi}), 
with the $m_h=0$ spectral function~(\ref{rho_phiphi_hphi}), leads to an analogous result for the parameter $S$:
\begin{eqnarray}
&&S-S^{(0)}_\epsilon 
\nonumber \\ 
&&\,=\, 
 - \frac{16\pi}{g^{2}\tan\theta_W} \!\int_{\epsilon\to 0}^{m_h^2} \frac{\Delta\rho_S(s)|_{h \varphi,\, m_h=0}}{s}\, ds  \! +\! \mO\left(m_h^2\right)
\nn\\
&&\, = \, -\, \frac{1}{12\pi} (\kappa_W^2-1)\log{\frac{\epsilon}{m_h^2}} \, +\, \mO\left(m_h^2\right) \, . 
\label{eq:S-Seps}
\end{eqnarray}

\section{Fermion form factors and spectral functions}

Let us consider the form factors for a generic spin--1 current $\mJ^\mu$ (which might be vector or axial-vector, custodial triplet or singlet) coupled to a fermion-antifermion pair in the final state~\cite{Nowakowski:2004cv}.  The corresponding matrix element has the general Lorentz decomposition:
 \begin{eqnarray}
\mathbb{F}^{\mJ\,\, \mu}_{\psi\bar{\psi}}
&\equiv & \langle \psi(p_1,\lambda_1)\overline{\psi}(p_2,\lambda_2)|\, \mJ^\mu |0\rangle 
\nonumber\\
&=&\bar{u}^{(\lambda_1)}(p_1)\bigg[ \gamma^\mu \, 
\mathcal{F}^{\mJ}_{1, \psi\bar{\psi}} (q^2)
+ \frac{i}{2}\sigma^{\mu\nu}q_\nu  \, 
\mathcal{F}^{\mJ}_{2, \psi\bar{\psi}} (q^2) \quad\;
\nonumber\\
&& 
+ \left( \gamma^\mu - \frac{2 m_\psi }{q^2} q^\mu \right) \gamma_5 \, 
\mathcal{F}^{\mJ}_{3, \psi\bar{\psi}} (q^2) 
\bigg] v^{(\lambda_2)}(p_2)\, ,
\end{eqnarray}
with $q=p_1+p_2$. 

We will make use of the optical theorem to relate these form factors to the spectral function of the $\mJ\mJ'$--correlator,
\begin{eqnarray}
\mbox{Im}\Pi^{\mu\nu}_{\mJ\mJ'}(q)   &=&  
  \left(-g^{\mu\nu}  + \frac{q^\mu q^\nu}{q^2} \right)\, \mbox{Im}\Pi_{\mJ\mJ'}(q^2),    
\nonumber 
\end{eqnarray} 
where unitarity provides the two-fermion absorptive cuts:
\begin{equation}
\mbox{Im}\Pi_{\mJ\mJ'}(q^2)= - \displaystyle{  \sum_{\psi}  \!\sum_{\lambda_1,\lambda_2}  } 
 \theta\!\left(q^2\!-\! 4m_\psi^2 \right) 
 \frac{ \beta_\psi  }{48\pi       } 
\mathbb{F}^{\mJ\,\, \mu}_{\psi\bar{\psi}}\, \mathbb{F}^{\mJ\,\,*}_{\psi\bar{\psi}\,\, \mu}, 
\end{equation} 
with the phase-space factor $\beta_\psi(q^2)=\sqrt{1-4m_\psi^2/q^2}$. After some algebra, one can extract the relation with the three possible form factors:
\begin{eqnarray}
&&\mbox{Im}\,\Pi_{\mJ\mJ'}(q^2) \,=\,  
\nonumber\\
&&\qquad  \displaystyle{  \sum_{\psi}    } 
 \theta\!\left(q^2\!-\! 4m_\psi^2 \right) 
\frac{ q^2 \beta_\psi  }{12\pi} 
\bigg[
\left( 1+\frac{2m_\psi^2}{q^2}\right) 
\mathcal{F}^{\mJ}_{1, \psi\bar{\psi}} \mathcal{F}^{\mJ'\,\, *}_{1, \psi\bar{\psi}}
+ \nonumber\\
&& \qquad 
+ \frac{q^2}{8}\left( 1+\frac{8m_\psi^2}{q^2}\right) 
\mathcal{F}^{\mJ}_{2, \psi\bar{\psi}} \mathcal{F}^{\mJ'\,\, *}_{2, \psi\bar{\psi}}
\nonumber
\end{eqnarray}
\begin{eqnarray}
&&\qquad 
+  \left( 1- \frac{4m_\psi^2}{q^2}\right)  
\mathcal{F}^{\mJ}_{3, \psi\bar{\psi}} \mathcal{F}^{\mJ'\,\, *}_{3, \psi\bar{\psi}}
\nonumber\\
&&\qquad 
+ \frac{3 m_\psi}{2} \bigg\{
\mathcal{F}^{\mJ}_{1, \psi\bar{\psi}} \mathcal{F}^{\mJ'\,\, *}_{2, \psi\bar{\psi}} + \mathcal{F}^{\mJ}_{2, \psi\bar{\psi}} \mathcal{F}^{\mJ'\,\, *}_{1, \psi\bar{\psi}}
\bigg\} 
\bigg]
\nonumber\\
&& \!=\!  \displaystyle{ \sum_{\psi}  }  
\frac{ \theta\!\left(q^2 \right) \!q^2  }{12\pi} \!
\bigg[ \!
\mathcal{F}^{\mJ}_{1, \psi\bar{\psi}} \mathcal{F}^{\mJ'\,\, *}_{1, \psi\bar{\psi}}
\!+\! \frac{q^2}{8}
\mathcal{F}^{\mJ}_{2, \psi\bar{\psi}} \mathcal{F}^{\mJ'\,\, *}_{2, \psi\bar{\psi}}
\!+\!
\mathcal{F}^{\mJ}_{3, \psi\bar{\psi}} \mathcal{F}^{\mJ'\,\, *}_{3, \psi\bar{\psi}}
\! \bigg] \nonumber\\
&&\qquad \!+\! \dots 
 \label{eq:unitarity-rel}
\end{eqnarray}
where the dots stand for $\mO(m_\psi)$ corrections, which vanish when the SM particle masses are neglected.

For the study of the parameter $S$ we will need the $W_3B$ correlator, where the corresponding currents are related to the singlet and triplet vector and axial-vector currents through, 
\begin{eqnarray}
\mJ_{W^3}^\mu &=& \,-\, \frac{g}{2} \left(\mathcal{V}_3^\mu - \mathcal{A}_3^\mu\right)\, , 
\nonumber\\
\mJ_{B}^\mu &=& \,-\, \frac{g'}{2} \left(\mathcal{V}_3^\mu + \mathcal{A}_3^\mu + 2 \mathcal{X}_{(0)}^\mu \right)\, , 
\end{eqnarray} 
stemming from the relations between the covariant sources and the physical gauge fields~\cite{lagrangian,lagrangian_color}:
\begin{eqnarray} 
v_\mu &=&v^a_\mu \frac{\sigma^a}{2} =
 \frac{r^a_\mu\!+\!\ell^a_\mu}{2}  \frac{\sigma^a}{2} =  \! \left( -\frac{1}{2}g'B_\mu \!-\! \frac{1}{2} g W_\mu^3 \right)\frac{\sigma^3}{2}+\dots
\nonumber\\
a_\mu &=& a^a_\mu \frac{\sigma^a}{2}=
 \frac{r^a_\mu\!-\!\ell^a_\mu}{2} 
\frac{\sigma^a}{2} =\! \left( -\frac{1}{2}g'B_\mu \!+\! \frac{1}{2} g W_\mu^3 \right)\frac{\sigma^3}{2} +\dots
\nonumber\\
X_\mu &=&  \, -\, g' B_\mu\, .
\end{eqnarray}  
Taking this into account the $W^3 B$ correlator is given by 
\begin{eqnarray}
\!\!\!\!\!\Pi_{30} 
&=& \frac{gg'}{4} \!\bigg[  
\Pi_{\mathcal{V}_3 \mathcal{V}_3 } \!-\! \Pi_{\mathcal{A}_3 \mathcal{A}_3} \!+\! 2 \Pi_{\mathcal{V}_3 \mathcal{X}_{(0)} }\! -\!2 \Pi_{\mathcal{A}_3 \mathcal{X}_{(0)}}   
\bigg] \!  .  
\label{eq:PiW3B}
\end{eqnarray}

Our resonance Lagrangian~(\ref{eq:Lagr})
produces the following form factors for a current $\mJ^\mu$ with a final $\psi\overline{\psi}$ state:
\begin{itemize}
    \item Form factors for a triplet vector current  
     ($\mJ^\mu=\mathcal{V}_3^\mu$):
\begin{eqnarray}
\mathcal{F}^{\mathcal{V}_3}_{1, \psi\bar{\psi}}
&=& T^3_\psi \, ,
\nonumber\\ 
\mathcal{F}^{\mathcal{V}_3}_{2, \psi\bar{\psi}}
&=&     \, -\,   4\sqrt{2} \, T^3_\psi\, 
\left( \frac{F_V C_0^{V_3^1}  }{M_{V}^2 - q^2} 
+\frac{\widetilde{F}_A \widetilde{C}_0^{A_3^1}  }{M_{A}^2 - q^2} \right)    \, , 
\nonumber \\ 
\mathcal{F}^{\mathcal{V}_3}_{3, \psi\bar{\psi}}
&=& 0  \, .  
\end{eqnarray}

    \item Form factors for a triplet axial-vector current 
     ($\mJ^\mu=\mathcal{A}_3^\mu$):
\begin{eqnarray} 
\mathcal{F}^{\mathcal{A}_3}_{1, \psi\bar{\psi}}
&=& 
 \frac{1}{\sqrt{2}}   
\frac{g' \widetilde{c}_{\mT} c^{\hat{V}_1}}{ M_{V_{1}  }^2-q^2 }
+
\frac{1}{\sqrt{2}}     
\frac{g' {c}_{\mT} \widetilde{c}^{\hat{A}_1}}{ M_{A_{1}  }^2-q^2 }\, ,
\nonumber\\ 
\mathcal{F}^{\mathcal{A}_3}_{2, \psi\bar{\psi}}
&=&       4\sqrt{2} \, T^3_\psi\, 
\left( \frac{\widetilde{F}_V C_0^{V_3^1}  }{M_{V}^2 - q^2} 
+\frac{F_A \widetilde{C}_0^{A_3^1}  }{M_{A}^2 - q^2} \right) \,,
\nonumber \\ 
\mathcal{F}^{\mathcal{A}_3}_{3, \psi\bar{\psi}}
&=&   T^3_\psi  \, + \,   
 \frac{1}{\sqrt{2}}    
\frac{g' \widetilde{c}_{\mT} \widetilde{c}^{\hat{V}_1}}{ M_{V_{1}  }^2-q^2 }  
+  
 \frac{1}{\sqrt{2}}     
\frac{g' {c}_{\mT} {c}^{\hat{A}_1}}{ M_{A_{1}  }^2-q^2 }  
\, .    
\end{eqnarray}
Note that $\mathcal{F}^{\mathcal{A}_3}_{1, \psi\bar{\psi}}$ and the non-SM part of $\mathcal{F}^{\mathcal{A}_3}_{3, \psi\bar{\psi}}$ are subleading in $g'$. Corrections of order ${g'}^{2}$ and higher are not shown here.
$M_R$ and $M_{R_1}$ denote the masses of the triplet $R$ and singlet $R_1$ resonances, respectively.

    \item Form factors for a singlet vector current 
     ($\mJ^\mu=\mathcal{X}_{(0)}^\mu$):
\begin{equation}
\mathcal{F}^{\mathcal{X}_{(0)}}_{1, \psi\bar{\psi}}
\,=\,  x_\psi  \, , \qquad \quad 
\mathcal{F}^{\mathcal{X}_{(0)}}_{2, \psi\bar{\psi}}
\,=\, \mathcal{F}^{\mathcal{X}_{(0)}}_{3, \psi\bar{\psi}} \,=\, 0 \, ,   
\end{equation}
where, as it has been explained previously, 
$x_\psi=\frac{1}{2}({\rm B-L})_\psi$ is the corresponding $U(1)_X$ charge of the fermion $\psi$
($1/6$ for quarks and $-1/2$ for leptons).

\end{itemize}   

In the SM limit 
all resonance couplings vanish and one has 
\begin{eqnarray}
\!\!\!\mathcal{F}^{\mathcal{V}_3}_{1, \psi\bar{\psi}} |^{\mathrm{SM}}   \!=\!  \mathcal{F}^{\mathcal{A}_3}_{3, \psi\bar{\psi}}|^{\mathrm{SM}}  \!=\!  T^3_\psi,\quad \mathcal{F}^{\mathcal{X}_{(0)}}_{1, \psi\bar{\psi}}|^{\mathrm{SM}}  \!=\!  x_\psi 
 ,  
\end{eqnarray}
with all the remaining form-factors vanishing.

For our calculation, we neglect the SM masses in the main text, that is, $\mathcal{O}(m_\psi)$ corrections are neglected in (\ref{eq:unitarity-rel}) and, consequently, only the last expression of (\ref{eq:unitarity-rel}) is considered. Moreover, we also neglect contributions that are subleading in $g'$, so that $\mathcal{F}^{\mathcal{A}_3}_{1, \psi\bar{\psi}}$ and the non-SM part of $\mathcal{F}^{\mathcal{A}_3}_{3, \psi\bar{\psi}}$ are discarded too.

\end{document}